\documentclass[a4paper,twocolumn]{revtex4}

\usepackage{amsmath,amssymb,epsfig}
\usepackage{graphicx}

\begin{document}

\title{DMFT+$\Sigma$ approach to disordered Hubbard model}

\author{E.\ Z.~Kuchinskii, M.\ V.~Sadovskii}
\affiliation{Institute for Electrophysics, Russian Academy of Sciences, 
Ural Branch,\\  Amundsen str. 106, Ekaterinburg 620016, Russia\\
M.N. Mikheev Institute for Metal Physics, Russian Academy of Sciences, 
Ural Branch,\\ S. Kovalevskaya str. 18, Ekaterinburg 620990, Russia}



\begin{abstract}

We briefly review the generalized dynamical mean-field theory DMFT+$\Sigma$
treatment of both repulsive and attractive disordered Hubbard models.
We examine the general problem of metal-insulator transition and  
the phase diagram in repulsive case, as well as BCS-BEC crossover region of 
attractive model, demonstrating certain universality of single -- electron 
properties under disordering in both models. We also discuss and compare  the 
results for the density of states and dynamic conductivity in both 
repulsive and attractive case and the generalized Anderson theorem behavior 
for superconducting critical temperature in disordered attractive case.
A brief discussion of Ginzburg -- Landau coefficients behavior under
disordering in BCS-BEC crossover region is also presented.


\end{abstract}

\maketitle

\newpage



\section{Introduction}

\label{intro}
Strongly correlated electronic systems, which are mainly realized in a range
of compounds containing transition or rare-earth elements with partially filled 
$3d$, $4f$ or $5f$ shells, attract attention of scientists because of 
their unusual physical properties and are notorious for major difficulties in
theoretical description. Perhaps the most significant development in this area 
was the discovery of high temperature superconductivity in copper oxides, which are 
considered to be the typical example of strongly correlated systems.

Early qualitative ideas formulated mainly by Mott \cite{NFM} as well as the introduction 
of the seminal Hubbard model \cite{Hubbard} inspired the hundreds of theoretical 
papers, which now constitute the separate branch of condensed matter theory. 
Probably the most impressive achievement if this field in recent years was 
the development of dynamical mean-field theory (DMFT), which provides an 
asymptotically exact solution for the Hubbard model in the limit of infinite 
dimensions \cite{MetzVoll89,vollha93,pruschke,georges96,Vollh10,PT}. 

Most of the studies of strongly correlated systems within Hubbard model are 
devoted to the case of repulsive interactions among electrons, which are
directly related to many topical problems, with most attention payed to 
the physics of high-T$_c$ superconductivity in cuprates and the general problem
of metal-insulator transition in cuprates and other similar oxides of transition
metals.

Another direction of research is the studies of the Hubbard model with attractive
interaction, which is related mainly to rather old problem of strong coupling 
superconductivity, especially to the theoretical description of the notorious BCS to 
BEC (Bardeen -- Cooper -- Schrieffer to Bose -- Einstein Condensation) crossover, 
which is also directly related to the problem of high-T$_c$ superconductivity in copper 
oxides. 
Starting with pioneering papers by Eagles and Leggett \cite{Eagles,Leggett} at $T=0$ and 
important progress achieved by Nozieres and Schmitt-Rink \cite{NS}, who suggested
an effective method to study the transition temperature crossover region, this field
has produced the large number of theoretical papers published during the recent years, 
including the successfull applications of DMFT approach.

This last area of research is also directly connected with recent progress in experimental 
studies of quantum gases in magnetic and optical dipole traps, as well as in optical lattices, 
with controllable parameters, such as density and interaction strength (cf. reviews \cite{BEC1,BEC2}), 
which has increased the interest to superconductivity (superfluidity of Fermions) with 
strong pairing interaction, including the region of BCS -- BEC crossover. 

In recent years we have developed the so called generalized DMFT+$\Sigma$ approach 
\cite{JTL05,PRB05,FNT06,UFN12}, which is very convenient for the studies of different 
additional interactions in repulsive Hubbard model, such as pseudogap fluctuations 
\cite{JTL05,PRB05,FNT06,UFN12}, disorder \cite{HubDis,HubDis2}, electron -- phonon 
interaction \cite{e_ph_DMFT}) etc. 
This approach is also well suited to analyze two--particle properties,
such as optical (dynamic) conductivity \cite{HubDis,PRB07}. 
In Ref. \cite{JETP14} we have used this approximation to calculate single -- particle properties of 
the normal phase and optical conductivity in attractive Hubbard model. 
Recently DMFT+$\Sigma$ approach was used by us to study disorder influence upon superconducting 
transition temperature in this model \cite{JTL14,JETP15}. 

Below we shall concentrate on discussion the of disorder effects in both repulsive and 
attractive Hubbard models. There are not so many works, devoted to the studies of 
disorder effects in Hubbard models, because of many theoretical complications, 
related to the problem of mutual interplay of disorder scattering and Hubbard interaction. 
We shall concentrate exclusively on our DMFT+$\Sigma$ approach, which is actually very 
convenient here and provides good interpolation scheme between different limiting cases. 
We shall discuss the results obtained in our previous work, similarities and dissimilarities 
of disorder effects in both repulsive and attractive Hubbard models, demonstrating in 
certain cases the universal dependences on disorder.

\section{The basics of DMFT+$\Sigma$ approach in disordered systems}

The Hamiltonian of disordered Hubbard model can be written as:
\begin{equation}
H=-t\sum_{\langle ij\rangle \sigma }a_{i\sigma }^{\dagger }a_{j\sigma
}+\sum_{i\sigma }\epsilon _{i}n_{i\sigma }+U\sum_{i}n_{i\uparrow
}n_{i\downarrow },  
\label{And_Hubb}
\end{equation}
where  $t>0$ is the transfer integral between nearest sites of the lattice, 
$U$ is the onsite interaction ($U>0$ in the case of repulsive interaction, while in 
the case of attraction $U<0$), $n_{i\sigma }=a_{i\sigma }^{\dagger }a_{i\sigma }^{{\phantom{\dagger}}}$ 
is the operator of the number of electrons on the lattice site $i$, $a_{i\sigma }$ 
($a_{i\sigma }^{\dagger}$) is annihilation (creation) operator for electron
with spin $\sigma$ on site $i$. The  local energy levels $\epsilon _{i}$ 
are assumed to be independent random variables at different lattice sites 
(Anderson disorder) \cite{Anderson58}.
To simplify diagram technique in the following we assume the Gaussian distribution
of these energy levels: 
\begin{equation}
\mathcal{P}(\epsilon _{i})=\frac{1}{\sqrt{2\pi}\Delta}\exp\left(
-\frac{\epsilon_{i}^2}{2\Delta^2}
\right)
\label{Gauss}
\end{equation}
Parameter $\Delta$ represents here the measure of disorder and this Gaussian 
random field (with ``white noise'' correlation on different lattice sites)
generates ``impurity'' scattering and leads to the standard diagram technique for
calculation of the ensemble averaged Green's functions \cite{Diagr}.

Generalized DMFT+$\Sigma$ approach \cite{JTL05,PRB05,FNT06,UFN12} 
extends the standard DMFT \cite{pruschke,georges96,Vollh10} introducing 
an additional self-energy  $\Sigma_{\bf p}(\varepsilon)$ 
(in general case momentum dependent), which is due to some interaction mechanism
outside the DMFT. It gives an effective procedure to calculate both single- and
two-particle properties \cite{HubDis,PRB07}. The single-particle Green's function 
is then written in the following form:
\begin{equation}
G(\varepsilon,{\bf p})=\frac{1}{\varepsilon+\mu-\varepsilon({\bf p})-\Sigma(\varepsilon)
-\Sigma_{\bf p}(\varepsilon)},
\label{Gk}
\end{equation}
where $\varepsilon({\bf p})$ is the ``bare'' electronic dispersion, 
while the total self-energy completely neglects the interference between 
the Hubbard and additional interaction and is given by the additive sum of the local 
self-energy $\Sigma (\varepsilon)$ of DMFT and ``external'' self-energy
$\Sigma_{\bf p}(\varepsilon)$. This conserves the standard structure of
DMFT equations \cite{pruschke,georges96,Vollh10}. However, there are
two important differences with standard DMFT. At each iteration of DMFT cycle
we recalculate the ``external'' self-energy $\Sigma_{\bf p}(\varepsilon)$ 
using some approximate scheme for the description of ``external'' interaction
and the local Green's function is ``dressed'' by $\Sigma_{\bf p}(\varepsilon)$ 
at each step of the standard DMFT procedure. 

\begin{figure}
\includegraphics[clip=true,width=0.45\textwidth]{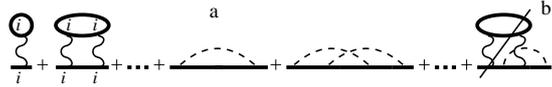}
\caption{Typical ``skeleton'' self-energy diagramms in
DMFT+$\Sigma$ approximation.}
\label{dDMFT_PG}
\end{figure}

In Fig. \ref{dDMFT_PG} we show the typical ``skeleton'' diagrams for self-energy
in DMFT+$\Sigma$. Here the first two terms are local DFMT self-energy diagrams
due to Hubbard interaction, while two diagrams in the middle show contributions 
to self-energy from additional  interaction (dashed interaction lines), and the 
last diagram (b) is a typical example of interference process which is neglected. 
Indeed, once we neglect such interference the total self-energy is defined as a 
simple sum of two contributions shown in Fig. \ref{dDMFT_PG}(a).

As an effective Anderson impurity solver in our DMFT calculations we have always 
used the numerical renormalization group \cite{NRGrev}, which allows to perform
calculations at pretty low temperatures.

For the self-energy due to disorder scattering produced by the Hamiltonian (\ref{And_Hubb}) 
we use below the simplest approximation neglecting the diagrams with
``intersecting'' interaction lines (like those in the fourth diagram of Fig. \ref{dDMFT_PG}(a)), 
i.e. the so called self-consistent Born approximation, represented by the third diagram in 
Fig. \ref{dDMFT_PG}(a). For the Gaussian distribution of site energies it is 
momentum independent and is given by:
\begin{equation}
\Sigma_{\bf p}(\varepsilon)\to\Delta^2\sum_{\bf p}G(\varepsilon,{\bf p}),
\label{BornSigma}
\end{equation}
where $G(\varepsilon,{\bf p})$ is the single-particle Green's function (\ref{Gk}), 
while $\Delta$ is the strength of site energy disorder.

In the following we shall consider mainly the three-dimensional system with
 ``bare'' semi -- elliptic density of states (per unit cell and one spin projection), 
with the total bandwidth $2D$, which is given by:
\begin{equation}
N_0(\varepsilon)=\frac{2}{\pi D^2}\sqrt{D^2-\varepsilon^2}.
\label{DOS13}
\end{equation}
In this case can directly demonstrate, that in DMFT+$\Sigma$ approximation
disorder influence upon single -- particle properties of disordered Hubbard model (both repulsive and
attractive) is completely described by effects of general band widening by disorder scattering.
Actually, in the system of self -- consistent equations DMFT+$\Sigma$ equations \cite{PRB05,UFN12,HubDis} both
the ``bare'' band spectrum and disorder scattering enter only on the stage of calculations of the local Green's 
function:
\begin{equation}
G_{ii}=\sum_{\bf p}G(\varepsilon,{\bf p}),
\label{Gii_det}
\end{equation}
where the full Green's function $G(\varepsilon,{\bf p})$ is determined by Eq. (\ref{Gk}), while
the self -- energy due to disorder, in self -- consistent Born approximation, is given by
Eq. (\ref{BornSigma}). Then, the local Green's function takes the following form:
\begin{eqnarray}
G_{ii}=\int_{-D}^{D}d\varepsilon' \frac{N_{0}(\varepsilon')}{\varepsilon+\mu-
\varepsilon'-\Sigma(\varepsilon)
-\Delta^2G_{ii}}=\nonumber\\
=\int_{-D}^{D}d\varepsilon' \frac{N_{0}(\varepsilon')}
{E_t-\varepsilon'},
\label{Gii_full}
\end{eqnarray}
where we have introduced $E_t=\varepsilon+\mu-\Sigma(\varepsilon)-\Delta^2G_{ii}$.
In the case of semi -- elliptic density of states (\ref{DOS13}) this integral can be
calculated in analytic form, so that the local Green's function reduces to: 
\begin{equation}
G_{ii}=2\frac{E_t-\sqrt{E_t^2-D^2}}{D^2}.
\label{Gii1}
\end{equation}
It can be easily seen that Eq. (\ref{Gii1}) represents one of the roots of quadratic equation:
\begin{equation}
G_{ii}^{-1}=E_t-\frac{D^2}{4}G_{ii},
\label{Gii_eq}
\end{equation}
reproducing the correct limit of $G_{ii}\to E_t^{-1}$ for infinitely narrow ($D\to 0$) band.
Then we can write:
\begin{eqnarray}
G_{ii}^{-1}=\varepsilon+\mu-\Sigma(\varepsilon)-\Delta^2G_{ii}-\frac{D^2}{4}G_{ii}=
\nonumber\\
=\varepsilon+\mu-\Sigma(\varepsilon)-\frac{D_{eff}^2}{4}G_{ii},
\label{Gii_eq_eff}
\end{eqnarray}
where we have introduced $D_{eff}$ -- an effective half-width of the band (in the absence of
electronic correlations, i.e. for $U=0$) widened by disorder scattering:
\begin{equation}
D_{eff}=D\sqrt{1+4\frac{\Delta^2}{D^2}}.
\label{Deff}
\end{equation}
Now comparing (\ref{Gii_full}), (\ref{Gii_eq}) and (\ref{Gii_eq_eff}), we immediately see, 
that the local Green's function can be written as: 
\begin{equation}
G_{ii}=\int_{-D_{eff}}^{D_{eff}}d\varepsilon' \frac{\tilde N_{0}(\varepsilon')}
{\varepsilon+\mu-\varepsilon'-\Sigma(\varepsilon)},
\label{Gii_full_eff}
\end{equation}
Here 
\begin{equation}
\tilde N_{0}(\varepsilon)=\frac{2}{\pi D_{eff}^2}\sqrt{D_{eff}^2-\varepsilon^2}
\label{DOSwidened}
\end{equation}
represents the density of states in the absence of interaction $U$
widened by disorder. The density of states in the presence of disorder remains 
semi -- elliptic, so that all effects of disorder scattering on single -- 
particle properties of disordered Hubbard model in DMFT+$\Sigma$ approximation 
are reduced only to disorder widening of conduction band, 
i.e. to the replacement $D\to D_{eff}$.

Within DMFT+$\Sigma$ approach we can also investigate the two-particle
properties \cite{HubDis,PRB07}. After the general analysis, based on 
Ward identity derived in Ref. \cite{PRB07}, we can show that
the real part of dynamical (optical) conductivity
in  DMFT+$\Sigma$ approximation is given by \cite{HubDis,PRB07}:
\begin{eqnarray}
{\rm{Re}}\sigma(\omega)=\frac{e^2\omega}{2\pi}
\int_{-\infty}^{\infty}d\varepsilon\left[f(\varepsilon_-)
-f(\varepsilon_+)\right]\times\nonumber\\
\times{\rm{Re}}\left\{\phi^{0RA}_{\varepsilon}(\omega)\left[1-
\frac{\Sigma^R(\varepsilon_+)-\Sigma^A(\varepsilon_-)}{\omega}\right]^2-
\right.\nonumber\\
\left.-\phi^{0RR}_{\varepsilon}(\omega)\left[1-
\frac{\Sigma^R(\varepsilon_+)-\Sigma^R(\varepsilon_-)}{\omega}\right]^2
\right\},\nonumber\\
\label{cond_final}
\end{eqnarray}
where $e$ is electronic charge, $f(\varepsilon_{\pm})$ --- Fermi distribution 
with $\varepsilon_{\pm}=\varepsilon\pm\frac{\omega}{2}$ and
\begin{eqnarray}
&&\phi^{0RR(RA)}_{\varepsilon}(\omega)=\nonumber\\
&&=\lim_{q\to 0}
\frac{\Phi^{0RR(RA)}_{\varepsilon}(\omega,{\bf q})-
\Phi^{0RR(RA)}_{\varepsilon}(\omega,0)}{q^2},\nonumber\\ 
\label{phi0}
\end{eqnarray}
where the two-particle loops (see details in Ref. \cite{HubDis}) 
$\Phi^{0RR(RA)}_{\varepsilon}(\omega,{\bf q})$ contain all vertex corrections 
from disorder scattering, but does not include any vertex corrections from 
Hubbard interaction. This considerably simplifies calculations of optical
conductivity within  DMFT+$\Sigma$ approximation, as we have only to solve
the single-particle problem determining the local self-energy
$\Sigma(\varepsilon_{\pm})$ via the DMFT+$\Sigma$ procedure, while 
non-trivial contributions from disorder scattering enter only via
$\Phi^{0RR(RA)}_{\varepsilon}(\omega,{\bf q})$, which can be calculated in some
appropriate approximation, neglecting vertex corrections from Hubbard interaction.
To be more specific, to obtain the loop contributions 
$\Phi^{0RR(RA)}_{\varepsilon}(\omega,{\bf q})$, determined by disorder
scattering, we can either use the usual ``ladder'' approximation for the case of
weak disorder, or following Ref. \cite{HubDis}, we can use the direct
generalization of the self-consistent theory of localization
\cite{VW,MS83,VW92}, which allows us to treat the case of strong enough 
disorder. In this approach conductivity is determined mainly by the generalized
diffusion coefficient obtained from simple extension of self-consistency
equation \cite{VW,MS83,VW92} of this theory, which is to be solved in
combination with DMFT+$\Sigma$ procedure \cite{HubDis}.

\section{Mott -- Anderson transition in disordered systems}

Below we present some of the most interesting results for repulsive 
Hubbard model at half-filling with semi -- elliptic bare density of states
(\ref{DOS13}) with the bandwidth $2D$ \cite{HubDis}, which qualitatively is well
suited to describe the three -- dimensional case.
Density of states below is given in units of number of states in
energy interval for cubic unit cell of the volume  $a^3$ 
($a$ is the lattice constant) and for one spin projection.
Conductivity values are always given in natural units of ${e^2}/{\hbar a}$.

\subsection{Evolution of the density of states}

In the standard DMFT approximation density of states of repulsive Hubbard model 
at half-filling has a typical three-peak structure \cite{georges96,pruschke,Bull}
with pretty narrow quasiparticle (central) peak at the Fermi level
and rather wide upper and lower Hubbard bands situated at energies
$\varepsilon\sim\pm U/2$.
As Hubbard repulsive interaction $U$ grows quasiparticle band narrows within 
the metallic phase and disappears at Mott-Hubbard metal-insulator transition 
at critical interaction value $U_{c2}/2D\approx 1.5$. At larger values of $U$ 
we observe insulating gap at the Fermi level.

\begin{figure}
\includegraphics[clip=true,width=0.45\textwidth]{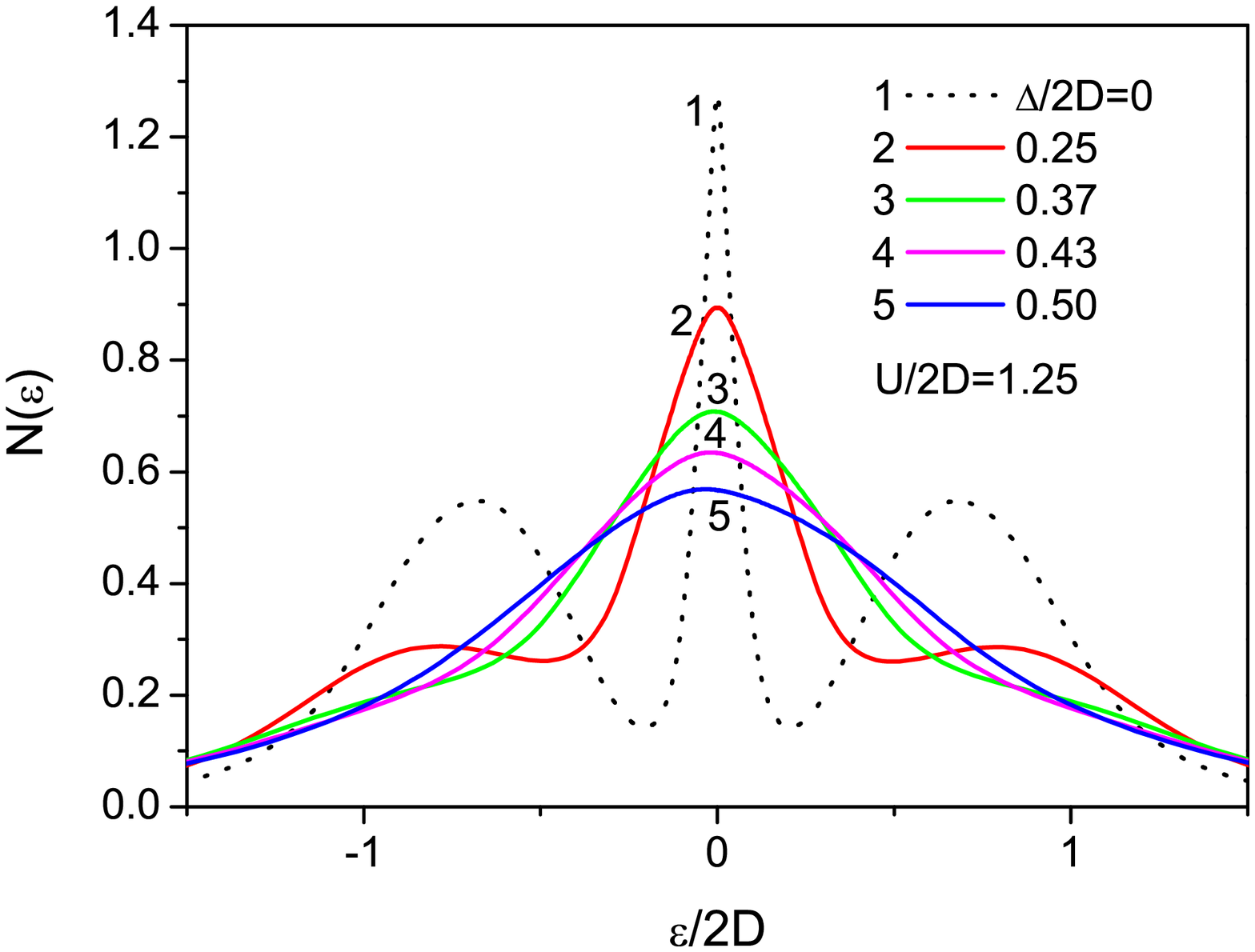}
\includegraphics[clip=true,width=0.45\textwidth]{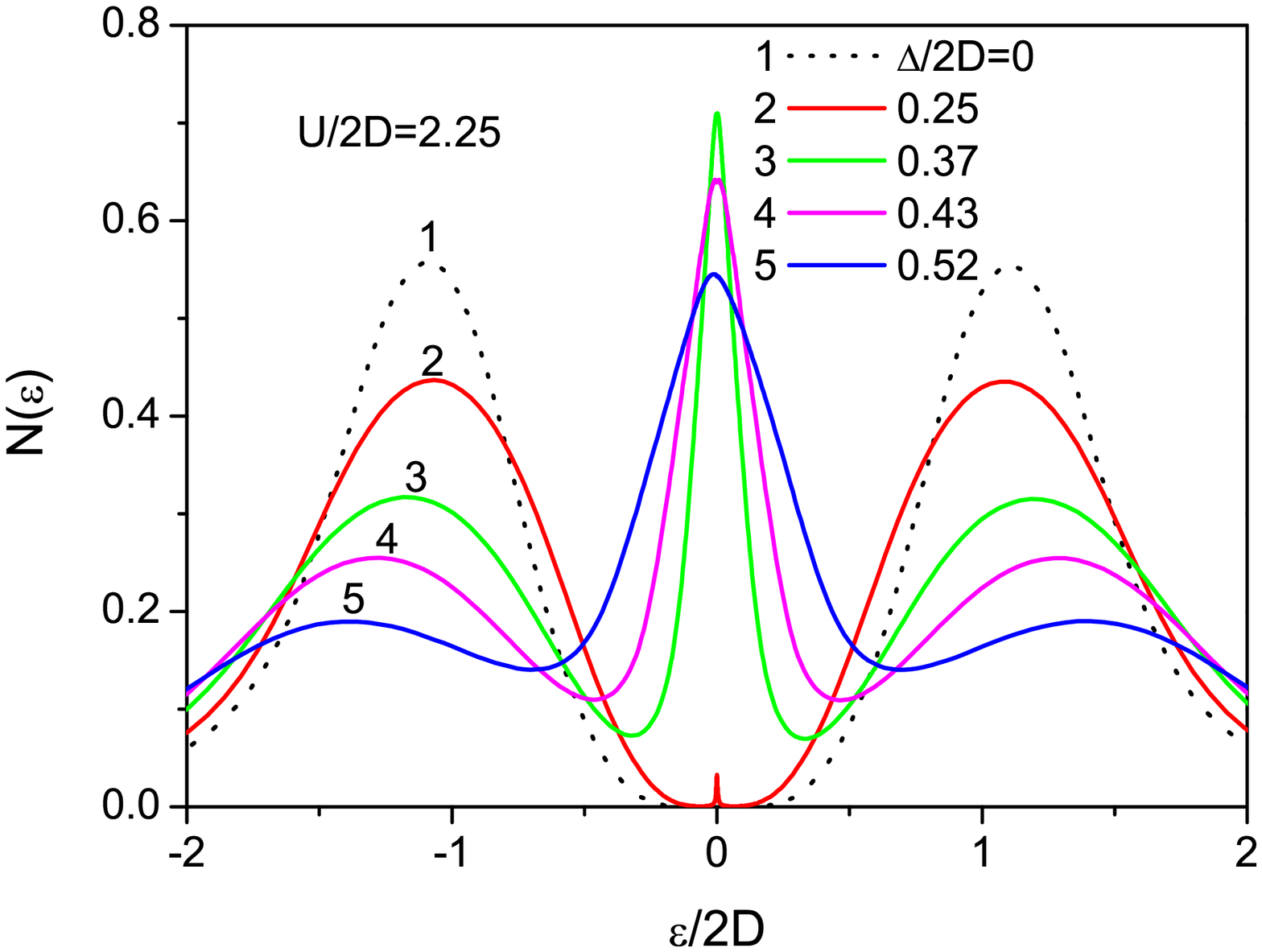}
\caption{Density of states of Hubbard model at half-filling
for different disorder levels $\Delta$ \cite{HubDis}.
(a) --- correlated metal with $U=2.5D$. 
(b) --- Mott insulator with $U=4.5D$. Temperature $T/2D$=0.0005.} 
\label{metDOS} 
\end{figure} 

In Fig. \ref{metDOS} we present our results \cite{HubDis} for DMFT+$\Sigma$ 
densities of states for typical strongly correlated metal with $U=2.5D$, 
both in the absense of disorder and for different values of disorder 
scattering $\Delta$, including strong enough values of disorder, which transforms 
correlated metal to correlated Anderson insulator. In metallic phase
disorder scattering leads to a typical broadening and suppression
of the density of states.

Much more unusual is the the result obtained for $U=4.5D$, typical for
Mott insulator phase and shown on Fig. \ref{metDOS}(b).
Here we observe the recovery of the central peak (quasiparticle band)
in the density of states with the increase of disorder, transforming Mott insulator
to correlated metal or to correlated Anderson insulator.
Similar density of states behavior for disordered Hubbard model was reported also in Ref. \cite{BV}, 
using direct numerical DMFT calculations in finite lattices.

\begin{figure}
\includegraphics[clip=true,width=0.45\textwidth]{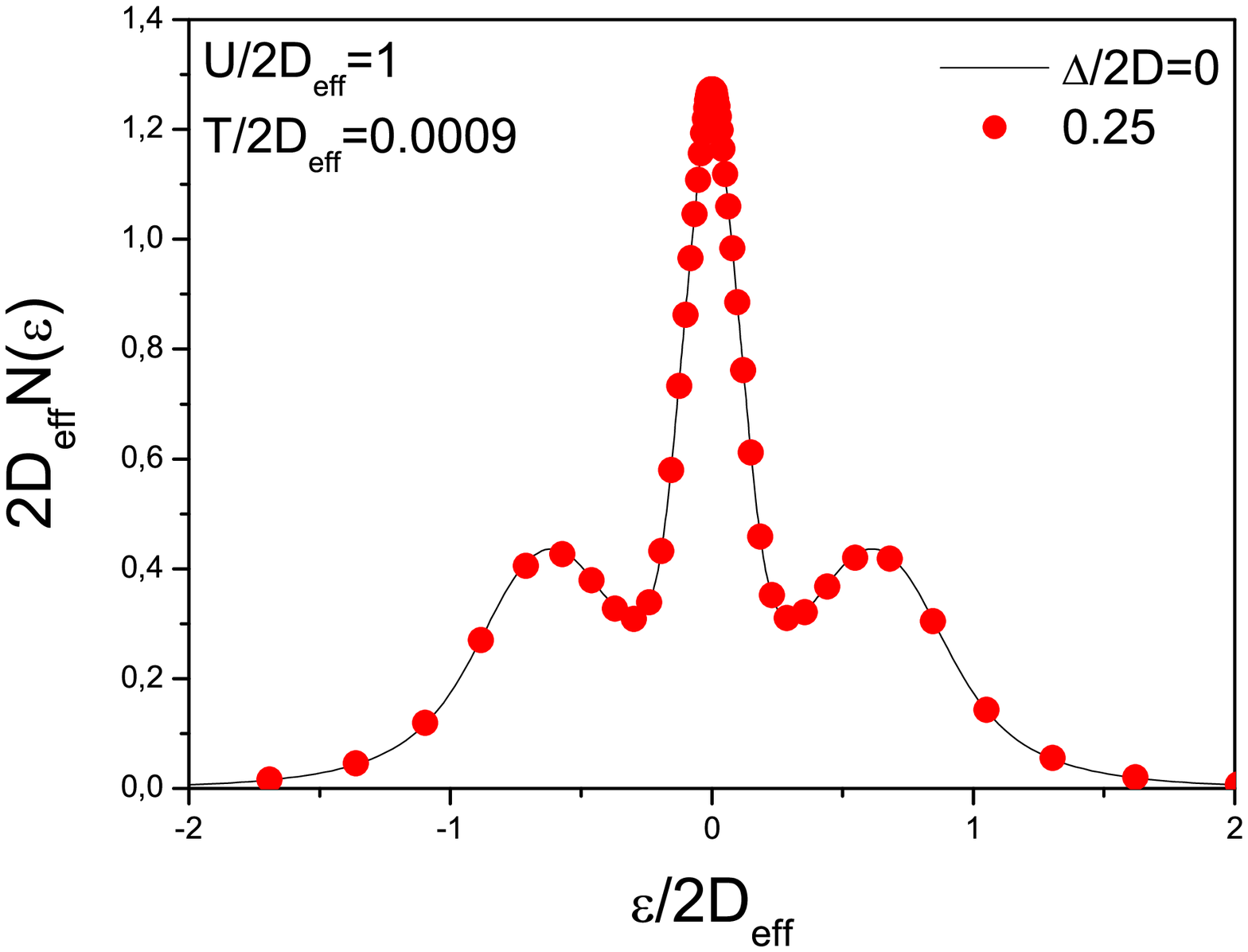}
\includegraphics[clip=true,width=0.45\textwidth]{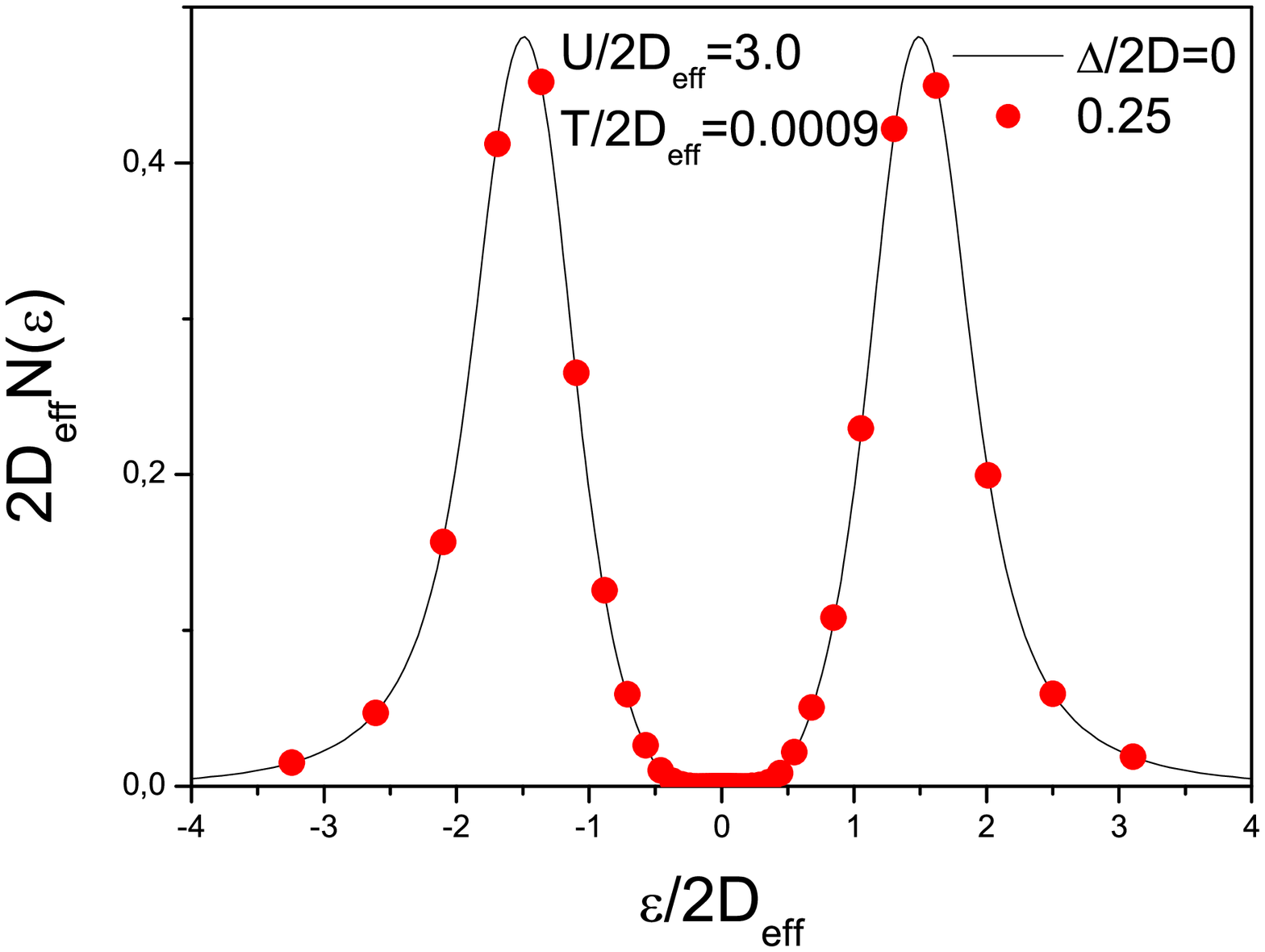}
\caption{\small Universal dependence of properly normalized density of states
on normalized energy $\varepsilon/2D_{eff}$ in Hubbard model for 
different disorder levels $\Delta$.
(a) --- correlated metal ($U/2D_{eff}$=1.0) with no disorder and for 
$\Delta/2D=$0.25.
(b) --- Mott insulator ($U/2D_{eff}$=3.0) without disorder and 
for $\Delta/2D=$0.25. Temperature $T/2D_{eff}$=0.0009.}

\label{uniDOS}
\end{figure}

Physical origin of such quite unexpected central peak restoration
is evident. Controlling parameter of metal-insulator transition
in DMFT is the ratio of Hubbard interaction $U$ to bare bandwidth $2D$.
Introduction of disorder (in the absense of Hubbard interaction) leads to new 
effective bandwidth $2D_{eff}$ (cf. (\ref{Deff})), growing with disorder. 
This leads to diminishing values of the ratio  $U/2D_{eff}$, which in its turn
causes restoration of the quasiparticle band.

More so, in complete accordance with analytic arguments presented above, the
density of states behavior in disordered Hubbard model with semi -- elliptic
actually demonstrates the universal dependence on disorder. This is clearly seen
from Fig. \ref{uniDOS}, where we show properly normalized typical densities of states
$2D_{eff}N(\varepsilon)$ in metallic (normalized interaction value $U/2D_{eff}$=1.0) 
and insulating phase (corresponding to $U/2D_{eff}$=3.0)  without disorder and
for the typical value of disorder scattering $\Delta/2D=$0.25. 
The densities of states in the absence and in the presence of disorder 
are actually described by the same (universal) dependences if expressed via 
properly normalized parameters.

In the absense of disorder one of the characteristic features of Mott-Hubbard 
metal-insulator transition is hysteresis behavior of the density of states, 
appearing with the decrease of $U$, starting from insulating phase \cite{georges96,Bull}.
Mott insulator phase remains (meta)stable down to rather small values of $U$ 
deep within the correlated metal phase and metallic phase is restored only at 
about $U_{c1}/2D\approx 1$. Corresponding interval of interaction parameter 
$U_{c1}<U<U_{c2}$ represents a coexistence region of metallic and Mott insulating 
phases, where, from a thermodynamic point of view, metallic phase is more stable
\cite{georges96,Bull,Blum}. Such hysteresis in density of states behavior 
is observed also in the presence of disorder \cite{HubDis,HubDis2}.

\subsection{Optical conductivity: Mott-Hubbard and Anderson transitions}

In the absence of disorder our calculations reproduce conventional DMFT results
\cite{pruschke,georges96}, with optical conductivity characterized
by typical Drude peak at low frequencies and wide maximum at about
$\omega\sim U$, which corresponds to optical transitions to the upper Hubbard band.
As $U$ grows Drude peak is suppressed and disappears completely at Mott transition.
Introduction of disorder leads to qualitative changes of the frequency dependence
of optical conductivity.

\begin{figure}
\includegraphics[clip=true,width=0.5\textwidth]{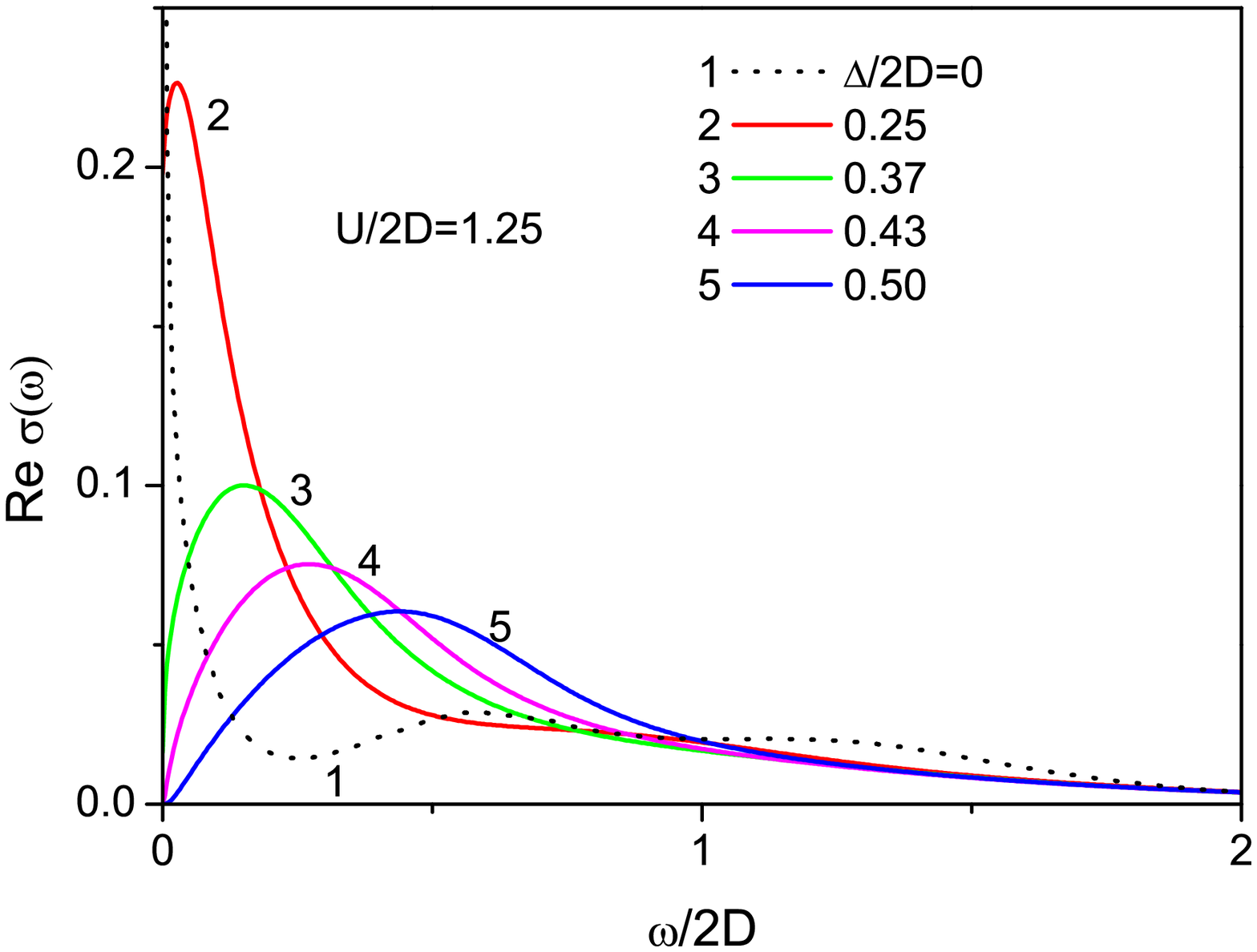}
\includegraphics[clip=true,width=0.42\textwidth]{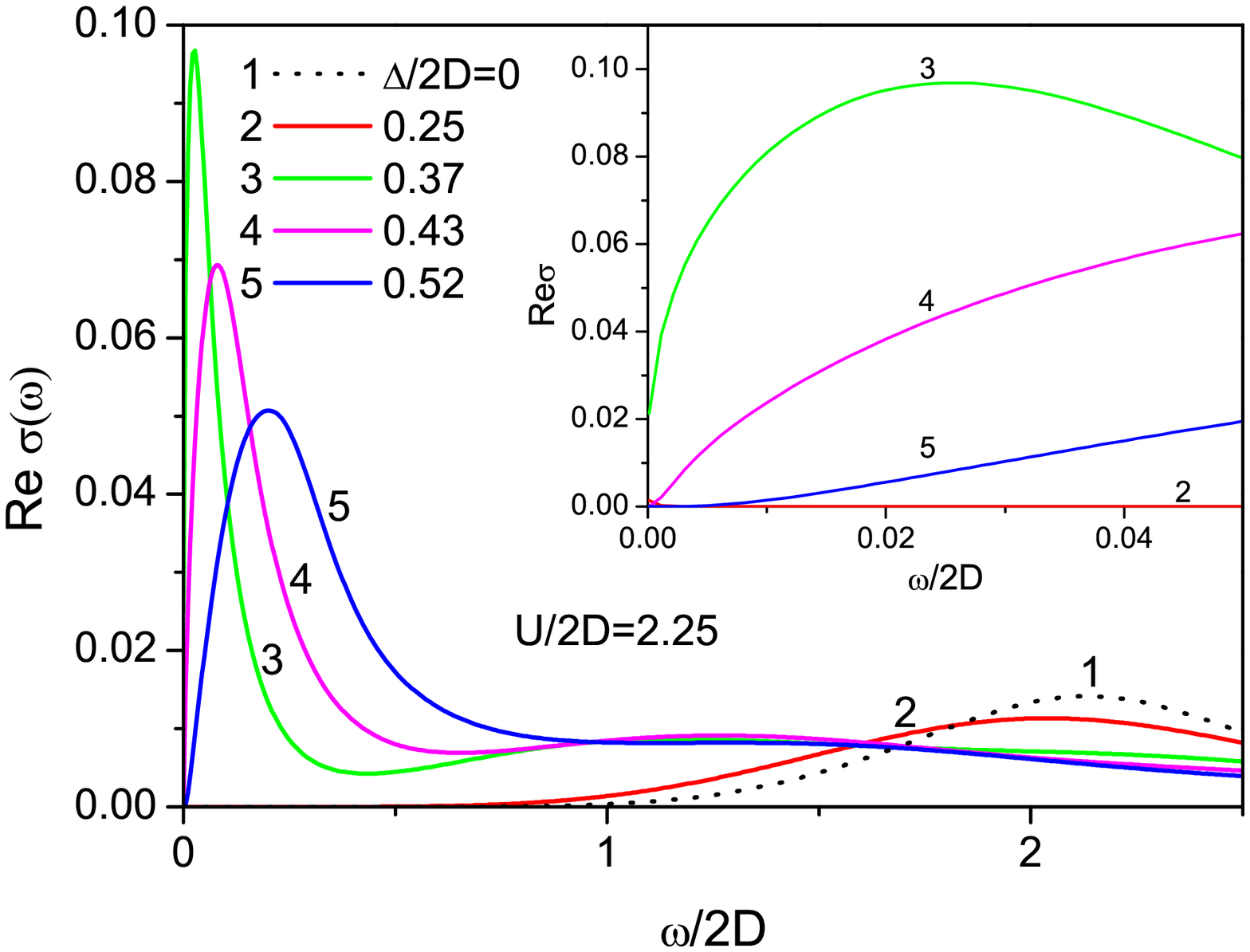}
\caption{\small Real part of optical conductivity of 
Hubbard model at half-filling for different
disorder levels $\Delta$ \cite{HubDis}.
(a) -- typical correlated metal with $U=2.5D$.
Curves 1,2, -- metallic phase, curve 3 corresponds to the mobility
edge (Anderson transition), curves 4,5 -- correlated Anderson insulator.
(b) -- typical Mott insulator with $U=4.5D$.
Curves 1,2 correspond to Mott insulator, curve 3 -- mobility edge
(Anderson transition), curves 4,5 -- correlated Anderson insulator.
Inset shows magnified low frequency region. Temperature $T/2D$=0.0005.} 
\label{met_cond} 
\end{figure}

Fig. \ref{met_cond}(a) shows the real part of optical 
conductivity of Hubbard model at half-filling for different disorder 
levels $\Delta$ and $U=2.5D$ typical for correlated metal. 
Transitions to the upper Hubbard bands at energies $\omega\sim U$
are almost unobservable. However it is clearly visible that
metallic Drude peak typically centered at zero frequency is broadened and 
suppressed by disorder, gradually transforming into a peak at finite frequency 
because of Anderson localization effects.
Anderson transition takes plase at $\Delta_c\approx 0.74D$
(corresponding to the curve 3 on all figures here).
Notice that this value explicitly depends on value of the cutoff in the equation
for the generalized diffusion coefficient, which is defined upto the coefficient 
of the order of unity \cite{MS83,Diagr}.
Naive expectations can lead to conclusion that narrow quasiparticle
band at the Fermi level (formed in a strongly correlated metal) may be 
localized much more easily than the usual conduction band. However,
these expectations are wrong and the band localizes only at rather large disorder
$\Delta_c\sim D$ of the order of conduction band width $\sim 2D$.
This is in qualitative agreement with the results for localization transition 
in  two-band model \cite{ErkS}.

In the DMFT+$\Sigma$ approach critical disorder value $\Delta_c$
does not depend on $U$ as interaction effects enter here
only through $\Delta\Sigma^{RA}(\omega)\to 0$  for $\omega\to 0$ 
(for $T=0$, $\varepsilon=0$), so that the influence of interaction 
at $\omega=0$ just disappears.
In fact this is the main shortcoming of DMFT+$\Sigma$ approach originating 
from the neglect of interference effect between interaction
and impurity scattering. Significant role of these interference effects is 
actually well known for a long time \cite{Lee85,ma}. However,
the neglect of these effects allows us to perform the
reasonable physical interpolation between two main limits -- that of 
Anderson transition because of disorder and Mott-Hubbard transition because of
strong correlations. 

On Fig. \ref{met_cond}(b) we show the real part of optical 
conductivity of Mott-Hubbard insulator with $U=4.5D$ for different disorder 
levels $\Delta$. In the inset we show low frequency data, demonstarting 
different types of conductivity behaviour, especially close to Anderson 
transition and within the Mott insulator phase. 
On the main part of the figure contribution to conductivity from transitions 
to upper Hubbard band at about $\omega\sim U$ is clearly seen. 
Disorder growth results in the appearance of finite conductivity for the frequencies
inside Mott-Hubbard gap, correlating with the restoration of quasiparticle
band in the density of states  within the gap as shown in Fig. \ref{metDOS}(b).
This conductivity for $\Delta<\Delta_c$ is metallic (finite in the static 
limit $\omega=0$), and for $\Delta>\Delta_c$ at low frequencies we get 
${\rm Re}\sigma(\omega)\sim\omega^2$, which is typical for Anderson insulator 
\cite{VW,Diagr,MS83,VW92}. 

A bit unusual is the appearance in ${\rm Re}\sigma(\omega)$ of a peak
at finite frequencies even in the metallic phase.
This happens because of importance of localization effects.
In the ``ladder'' approximation for $\Phi^{0RA}_{\varepsilon}(\omega,{\bf q})$
which neglects all localization effects we obtain the usual Drude peak
at $\omega=0$  \cite{HubDis}, while taking into account localization effects
shifts the peak in ${\rm Re}\sigma(\omega)$ to finite frequencies. 

Above we presented the data for conductivity data obtained for the case
of increase of $U$ from metallic to Mott insulator phase.
As $U$ decreases from Mott insulator phase we observe hysteresis of conductivity
in coexistence region defined (in the absense of disorder) by inequality
$U_{c1}<U<U_{c2}$. Hysteresis of conductivity is also observed in the 
coexistence region in the presence of disorder. More details can be found
in Refs. \cite{HubDis,HubDis2}.

In general, the picture of conductivity behavior obtained in DMFT+$\Sigma$
approximation is rather rich, demonstarting both Mott -- Hubbard transition
due to strong correlations and disorder induced Anderson (localization)
transition. The complicate behavior under disordering is essentially determined 
by two -- particle Green's function behavior and does not shown a kind of
universality, demonstrated above for the single -- particle density of states.

\subsection{Phase diagram of disordered Hubbard model at half-filling}

Phase diagram of repulsive disordered Hubbard model at half-filling was studied
in Ref. \cite{BV}, using direct DMFT numerics for lattices with finite 
number of sites with random realizations of energies $\epsilon_i$ in 
(\ref{And_Hubb}), with subsequent averaging over many lattice realizations to obtain
the averaged density of states and geometric mean local density of states, which 
allows to determine the critical disorder for Anderson transition.
Below we present our results on disordered Hubbard model phase diagram
obtained from density of states and optical conductivity calculations in  
DMFT+$\Sigma$ approach \cite{HubDis}. 

\begin{figure}
\includegraphics[clip=true,width=0.5\textwidth]{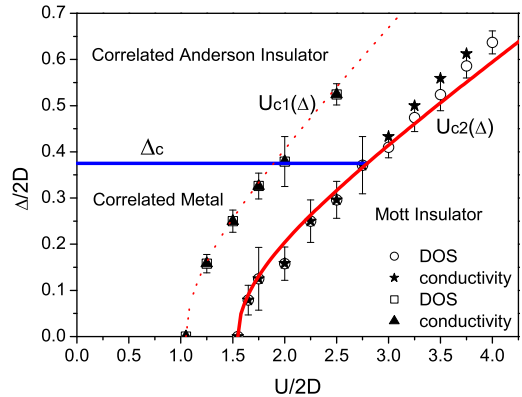}
\caption{
Phase diagram of disordered Hubbard model \cite{HubDis}.
Continuous curves are Mott insulator phase boundaries $U_{c1,c2}(\Delta)$
obtained from analytical estimate of Eq. (\ref{Uc}), different symbols
represent results for these boudaries obtained from calculations from density
of states  and optical conductivity. 
Line of Anderson transition is given by $\Delta_c=0.37$.}
\label{ph_diag} 
\end{figure} 

Calculated disorder -- correlation strength $(\Delta,U)$ phase diagram at 
zero temperature is shown in  Fig. \ref{ph_diag} (actually calculations 
were performed at very low $T/2D$=0.0005).
Anderson transition line $\Delta_c\approx 0.74D$ is defined
as disorder strength for which static conductivity becomes zero at  $T=0$.
Mott-Hubbard transition can be detected either from central peak
disappearance in the density of states or from optical conductivity by observation of
gap closing in the insulating phase or from static conductivity disappearance
in the metallic phase.

We have already noticed that DMFT+$\Sigma$ approximation gives
universal ($U$ independent) value of critical disorder $\Delta_c$
because of neglect of interference between disorder scattering
and Hubbard interaction. This leads to the difference
between the phase diagram of Fig.  \ref{ph_diag} and the one obtained 
by numerical simulations in Ref. \cite{BV}. At the same time 
the qualitative form of our phase diagram is highly nontrivial and 
qualitatively coincide with results of Ref. \cite{BV}.
Main difference is conservation of Hubbard bands in our results even in the 
limit of high enough disorder, while in the Ref. \cite{BV} they just disappear.
Phase coexistence region in Fig. \ref{ph_diag} 
slowly widens with disorder growth instead of vanishing
at some ``critical'' point as on phase diagram of Ref. \cite{BV}.
Coexistence boundaries (Mott insulator phase boundaries), obtained with 
decrease or increase of $U$, represented by curves $U_{c1}(\Delta)$ 
and $U_{c2}(\Delta)$ on Fig. \ref{ph_diag}, can actually be obtained from 
the simple equation:

\begin{equation}
\frac{U_{c1,c2}(\Delta)}{D_{eff}}=\frac{U_{c1,c2}}{D},
\label{UcW13}
\end{equation}
where effective bandwidth in the presence of disorder is calculated
for $U=0$ within self-consistent Born approximation (\ref{BornSigma}),
(\ref{Deff}). Thus the boundaries of coexistence region (which define also
the boundaries of Mott insulator phase) are given by:
\begin{equation}
U_{c1,c2}(\Delta)=U_{c1,c2}\sqrt{1+4\frac{\Delta^2}{D^2}}
\label{Uc}
\end{equation}
which are shown in Fig. \ref{ph_diag} by dotted and solid lines.
Phase transition points detected from disappearance of quasiparticle peak
as well as points following from qualitative changes of conductivity
behaviour are shown in Fig. \ref{ph_diag} by different symbols.
These symbols demonstrate very good agreement with analytical results
confirming the choice of the ratio (\ref{UcW13}) as a controlling parameter 
of Mott transition in the presence of disorder.
Thus, this transition is essentially controlled by simple band -- widening
effects due to disorder scattering, similarly to density of states behavior
demonstrated above.

Note that the values of  normalized density of states $2D_{eff}N(\varepsilon)$ 
are universal along each of these boundaries, as well 
as along any curve in ($\Delta$,$U$) -- plane, determined by the equation:
\begin{equation}
U(\Delta)=U(0)\sqrt{1+4\frac{\Delta^2}{D^2}}
\label{Ucc}
\end{equation}
in accordance with our discussion on the universal dependence of the densities
of states on disorder presented above.

Essentially similar results were obtained for the density of states behavior,
dynamic conductivity and phase diagram \cite{HubDis2} in the case of the 
conduction band with ``flat'' density of states in the absence of disorder
and interactions, which qualitatively corresponds to the two -- dimensional case.
This is not surprising, as both large enough disorder and interactions
transform the ``flat'' band into a kind of smeared semi -- elliptic band. Some
explicit examples of this kind of behavior will be presented below for the 
case of attractive Hubbard model.

\section{Attractive Hubbard model with disorder}

The studies of superconductivity in BCS -- BEC crossover region attracts theorists
for rather long time \cite{Leggett} and most important advance here was made
by Nozieres and Schmitt-Rink \cite{NS}, who proposed an effective approach to
describe $T_c$ crossover. Attractive Hubbard model is probably the simplest
model allowing theoretical studies of BCS-BEC crossover \cite{NS}. 
This model was studied within DMFT in a number of recent papers
\cite{Keller01,Toschi04,Bauer09,Koga11}. However only few results were
obtained for the normal (non-superconducting) phase of this model, especially in
disordered case. Similarly, there were practically no studies of two -- particle 
properties, such as optical conductivity. Below we present a summary of our
results obtained within DMFT+$\Sigma$ approach and make comparison with similar
results for repulsive Hubbard model.

\subsection{Density of states and optical conductivity}

\begin{figure}
\includegraphics[clip=true,width=0.5\textwidth]{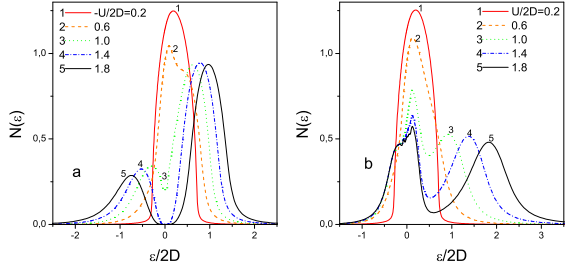}
\caption{Densities of states and for different 
values of Hubbard attraction (a) and repulsion (b). 
Temperature $T/2D=0.05$.}
\label{fig1}
\end{figure}

In the special case of half -- filled band ($n=1$) the densities of states of attractive 
and repulsive Hubbard models just coincide (due to exact mapping of these models onto 
each other). Thus, below we discuss the more typical case of quarter -- filled band ($n=0.5$).
In Fig. \ref{fig1} we show densities of states obtained for $T/2D=0.05$ for different values 
of attractive interaction ($U<0$). Fig. \ref{fig1}(a) should be compared with Fig. \ref{fig1}(b), 
where we present similar results for repulsive ($U>0$) case. 
We can see that the densities of states close to the Fermi
level drop with the growth of $U$, both for attraction (Fig. \ref{fig1}(a)) and 
repulsion (Fig. \ref{fig1}(b)), but significant growth of $|U|$ in repulsive case leads only
to vanishing quasiparticle peak, so that the density of states at the Fermi level becomes practically
independent of $U$, while in attractive case the growth of $|U|$ leads to superconducting
pseudogap opening at the Fermi level (curves 5, 6 in Fig. \ref{fig1}(a)) and for $|U|/2D>1.2$ 
we observe the full gap opening at the Fermi level (curves 7-9 in Fig. \ref{fig1}(a)). 
This gap is not directly related to the emergence of superconducting state, but is due to the
appearance of preformed Cooper pairs at the temperatures larger than superconducting transition 
temperature (which is lower, than the temperature $T/2D=0.05$ used in our calculations). 
Here we actually observe the important difference between attractive and repulsive cases --- 
in case of repulsion deviation from half -- filling leads to metallic state for arbitrary 
values of $U$ and insulating gap at large $U$ opens not at the Fermi level.

\begin{figure}
\includegraphics[clip=true,width=0.5\textwidth]{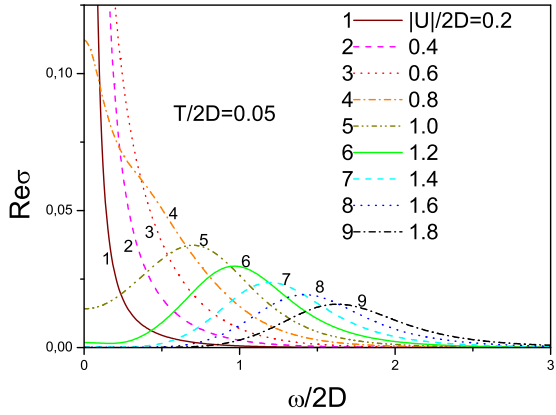}
\caption{Optical conductivity for different values of attractive Hubbard attraction. 
Temperature $T/2D=0.05$.}
\label{fig_2}
\end{figure}

This picture of density of states evolution with the growth of $|U|$ is also supported by the
behavior of optical conductivity shown in Fig. \ref{fig_2}. We see that the growth of $|U|$ 
leads to the replacement of Drude peak at zero frequency (curves 1-3 in Fig. \ref{fig_2}) by
pseudogap dip (curves 5, 6 in Fig. \ref{fig_2}) and wide maximum of conductivity at finite
frequency, connected with transitions across the pseudogap. The further increase of $|U|$
leads to opening of the full gap in optical conductivity due to formation of Cooper 
pairs (curves 7-9 in Fig. \ref{fig_2}).

\begin{figure}
\newpage
\includegraphics[clip=true,width=0.5\textwidth]{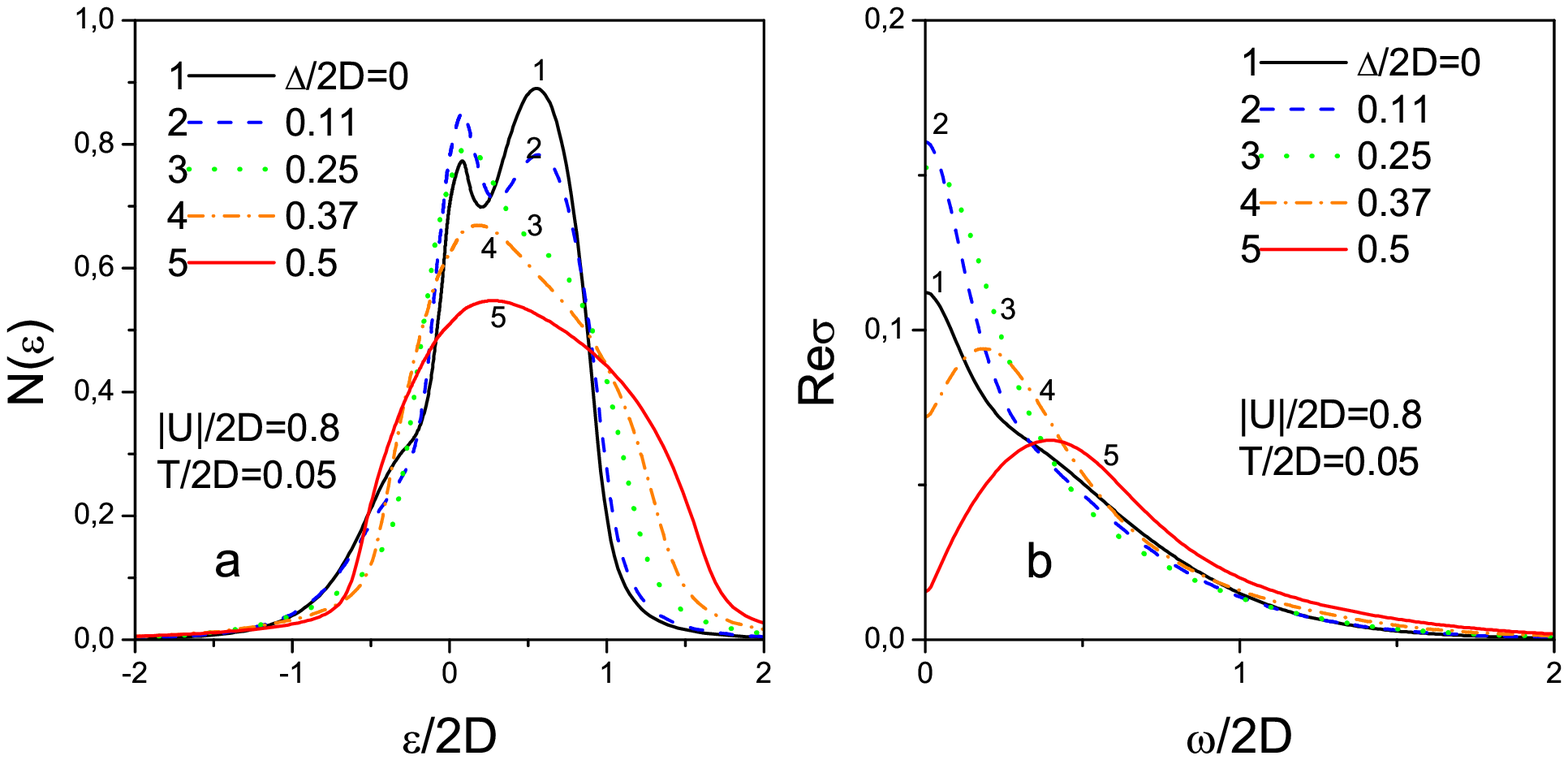}
\includegraphics[clip=true,width=0.5\textwidth]{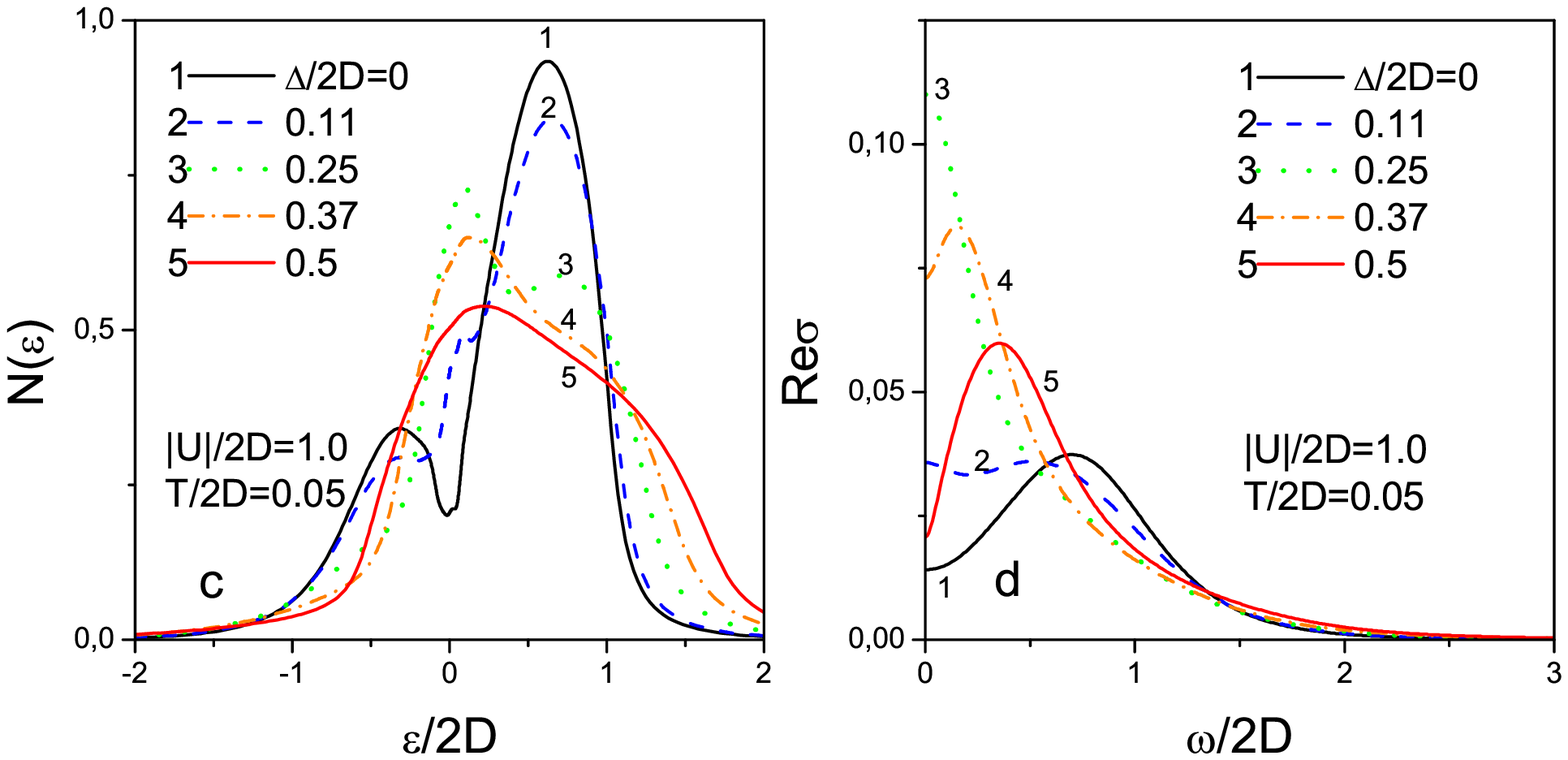}
\includegraphics[clip=true,width=0.5\textwidth]{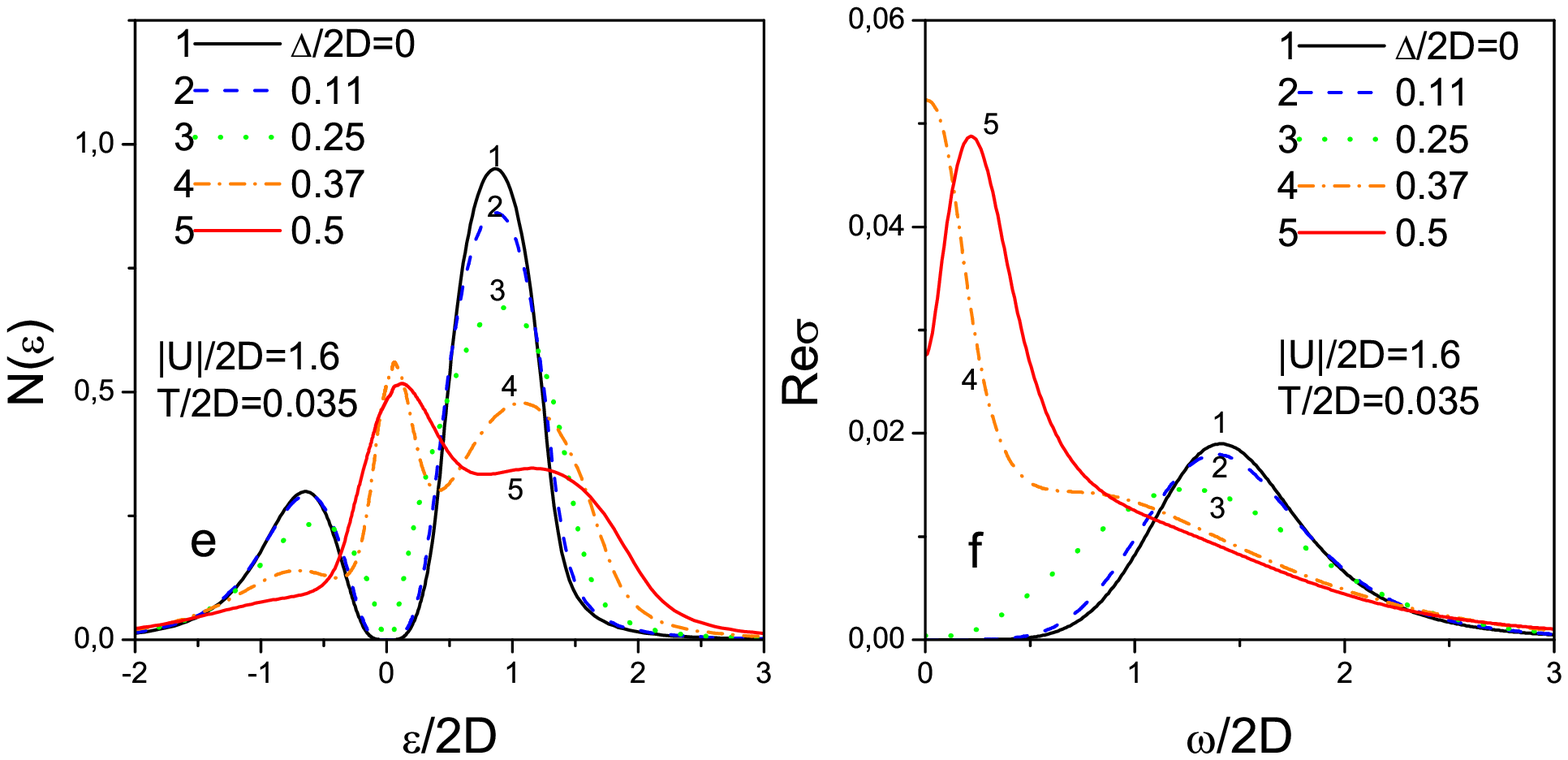}
\caption{Evolution of the density of states (left panels) and optical
conductivity (right panels) with disorder for different values of
$U$ ($|U|/2D=0.8$ - (a),(b); $|U|/2D=1$ - (c),(d); $|U|/2D=1.6$ - (e),(f)).}
\label{fig6}
\end{figure}

In Fig. \ref{fig6} we present the evolution of the density of states and optical conductivity 
with changing disorder. At weak enough attraction ($|U|/2D=0.8$, Fig. \ref{fig6}(a),(b)), 
the growth of disorder just widens the density of states.
Disorder effectively masks peculiarities of the density of states due to correlation effects.
In particular, quasiparticle peak and the ``wings'' due to upper and lower Hubbard bands 
present in Fig. \ref{fig6}(a) in the absence of disorder completely vanish at
strong enough disorder. 
Evolution of optical conductivity with the growth of disorder $\Delta$, shown in 
Fig. \ref{fig6}(b), is in general agreement with the evolution of density of states. 
Weak enough disordering (curves 1-3 in Fig. \ref{fig6}(b)), leads to some growth of static 
conductivity, which is connected with suppression of correlation effects at the Fermi level
(curves 1-3 in Fig. \ref{fig6}(a). The further growth of disorder leads to significant
widening of the band and the drop of the density of states (curve 5 in Fig. \ref{fig6}(a),(b)), 
which leads to the drop of static conductivity. Finally, the growth of disorder leads to
Anderson localization which takes place at $\Delta/2D=0.37$ for $T=0$ \cite{HubDis}. 
However, here we consider the case of high enough temperature $T/2D=0.05$, so that static 
conductivity (see curves 6, 7 in Fig. \ref{fig6}(b)) always remains finite, though the
localization behavior is also clearly seen and $\sigma(\omega)\sim\omega^2$. 
At larger value of attractive interaction $|U|/2D=1$ the evolution of the density of states 
and optical conductivity is more or less similar (Fig. \ref{fig6}(c),(d) ). 
However, in the absence of disorder we observe here Cooper pairing
pseudogap in the density of states, while disorder leads to its suppression, leading both to the 
growth of the density of states at the Fermi level and related growth of static conductivity.
Finally, at still larger attraction $|U|/2D=1.6$ (Fig. \ref{fig6}(e),(f)) in the absence of 
disorder there is the real Cooper pairing gap in the density of states. This gap is also evident
in optical conductivity. With the growth of disorder Cooper pairing gap both in the density
of states and conductivity becomes narrower (curves 2-5). Further growth of disorder leads
to complete suppression of this gap and restoration of metallic state with finite density
of states at the Fermi level and finite static conductivity. This closure of Cooper gap is
obviously related to the effective growth of the conduction bandwidth $2D_{eff}$, which leads to 
the lowering of $|U|/2D_{eff}$ ratio, which actually controls the formation of Cooper gap. 
Situation here is similar to the closure of Mott gap by disorder in repulsive Hubbard 
model discussed above \cite{HubDis}.
However, at large disorder (curve 7 in Fig. \ref{fig6}(f)) we clearly observe localization
behavior, so that the growth of disorder at $T=0$ will first lead to metallic state (the
closure of Cooper pairing gap), while the further growth of disorder will induce Anderson 
metal -- insulator transition. Similar picture is observed for large positive $U$ at half-filling
($n=1$) \cite{HubDis}, where the growth of disorder leads to Mott insulator -- correlated metal 
-- Anderson insulator transition. 

\begin{figure}
\includegraphics[clip=true,width=0.5\textwidth]{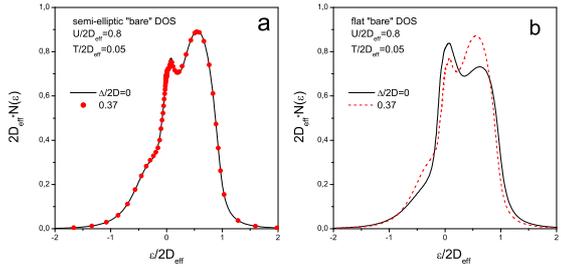}
\caption{Universal dependence of the density of states on disorder
in attractive Hubbard model: 
(a) ---  the model of semi -- elliptic ``bare'' density of states.
(b) --- the model of flat ``bare'' density of states.}
\label{fig2}
\end{figure}

Let us now demonstrate the universality of disorder dependence of the density of states
as an example of the most important single -- particle property. 
Let us concentrate on the most typical case of the density of states evolution shown in 
Fig. \ref{fig6} (a). We can easily convince ourselves, that this evolution is only due to
the general widening of the band due to disorder (cf. (\ref{Deff})), as all the data for 
the density of states fit the same universal curve replotted in appropriate new 
variables, with all energies (and temperature) normalized 
by the effective bandwidth by replacing $D\rightarrow D_{eff}$, as shown in 
Fig. \ref{fig2}(a), in complete accordance with results obtained above in
repulsive Hubbard model for semi -- ellipric band.

In the case of initial (``bare'') conduction band with flat density of states, there is no
complete universality, as is seen from Fig. \ref{fig2}(b) for low enough values of
disorder. However, for large enough disorders the dashed curve in 
Fig. \ref{fig2}(b) practically coincides with universal curve for the density of
states shown in Fig. \ref{fig2}(a). This reflects the simple fact, that at large 
enough disorders the flat density of states is effectively transformed into 
semi -- elliptic one\cite{JETP15}.

\subsection{Generalized Anderson theorem}

Superconducting transition temperature $T_c$ in general is not a single -- particle
characteristic of the system. Cooper instability, determining $T_c$ is 
related to divergence of two -- particle loop in Cooper channel. In the weak
coupling limit, when superconductivity is due to the appearance of Cooper pairs 
at $T_c$, disorder only slightly influences superconductivity with
$s$-wave pairing \cite{SCLoc,Genn}. This is the essence of the so called 
Anderson theorem and changes of $T_c$ are due only to the relatively 
small changes of the density of states at the Fermi level induced by disorder. 

In region of BCS -- BEC crossover and in the strong coupling region 
Nozieres -- Schmitt-Rink approach \cite{NS} assumes, that corrections
due to strong pairing attraction significantly change the chemical potential of
the system, while possible correction due to this interaction to Cooper
instability condition can be neglected, so that we can always use the weak
coupling (ladder) approximation. Then the condition of Cooper
instability in disordered Hubbard model takes the form:
\begin{equation}
1=-|U|\chi_0(q=0,\omega_m=0)
\label{Cupper}
\end{equation}
where
\begin{equation}
\chi_0(q=0,\omega_m=0)=-T\sum_{n}\sum_{\bf pp'}\Phi_{\bf pp'}(\varepsilon_n)
\label{chi}
\end{equation}
represents the two -- particle loop (susceptibility) in Cooper channel ``dressed'' only by 
disorder scattering, and $\Phi_{\bf pp'}(\varepsilon_n)$ is the averaged two -- particle 
Green's function in Cooper channel  ($\omega_m=2\pi mT$ and $\varepsilon_n=\pi T(2n+1)$ 
are the usual Boson and Fermion Matsubara frequencies).

\begin{figure}
\includegraphics[clip=true,width=0.45\textwidth]{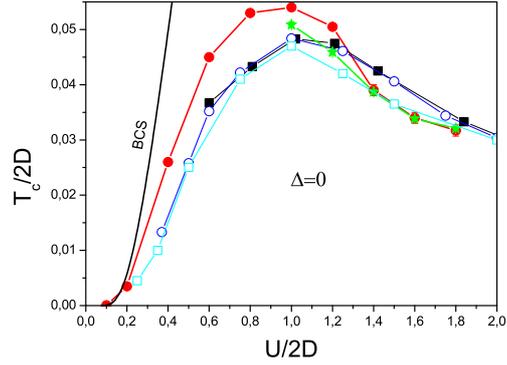}
\caption{Dependence of superconducting critical temperature on attractive
interaction strength. Black squares, white circles and white squares show the
results of Refs. \cite{Keller01},\cite{Toschi04},\cite{Koga11}
respectively for quarter-filled band with $n=0.5$. Stars represent the results
obtained numerically from the criterion of instability of the normal phase in Ref. \cite{JETP14}.
Filled circles show $T_c$ in Nozieres -- Schmitt-Rink approximations, combined
with DMFT \cite{JETP14}. Continuous black curve represents the result of BCS theory.}
\label{fig10a}
\end{figure}

Using the exact Ward identity, derived in Ref. \cite{PRB07} 
\begin{eqnarray}
&&G(\varepsilon_n,{\bf p})-G(-\varepsilon_n,-{\bf p})=\nonumber\\
&&=-\sum_{\bf p'}\Phi_{\bf pp'}(\varepsilon_n)(G_0^{-1}(\varepsilon_n,{\bf p'})-
G_0^{-1}(-\varepsilon_n,-{\bf p'})),\nonumber\\
\label{Word}
\end{eqnarray}
where $G(\varepsilon_n,{\bf p})$ is the impurity averaged (but not containing Hubbard interaction corrections!) 
single -- particle Green's function, we can show \cite{JETP15}
that Cooper susceptibility (\ref{chi}) is given by:
\begin{eqnarray}
&&\chi_0(q=0,\omega_m=0)=\nonumber\\
&&=T\sum_{n}\frac{\sum_{\bf p}G(\varepsilon_n,{\bf p})-\sum_{\bf p}G(-\varepsilon_n,{\bf p})}{2i\varepsilon_n}=
\nonumber\\
&&=T\sum_{n}\frac{\sum_{\bf p}G(\varepsilon_n,{\bf p})}{i\varepsilon_n}.
\label{chi1}
\end{eqnarray}
After the standard summation over Matsubara frequencies \cite{Diagr} we get:
\begin{eqnarray}
&&\chi_0(q=0,\omega_m=0)=\nonumber\\
&&=\frac{1}{4\pi i}\int_{-\infty}^{\infty}d\varepsilon
\frac{\sum_{\bf p}G^R(\varepsilon,{\bf p})-\sum_{\bf p}G^A(\varepsilon,{\bf p})}{\varepsilon}th\frac{\varepsilon}{2T}\nonumber\\
&&=-\int_{-\infty}^{\infty}d\varepsilon\frac{\tilde N_0(\varepsilon)}{2\varepsilon}th\frac{\varepsilon}{2T},
\label{chi2}
\end{eqnarray}
where $\tilde N_0(\varepsilon)$ is the density of states ($U=0$)  renormalized by 
disorder scattering. In Eq. (\ref{chi2}) the energy $\varepsilon$ origin is at 
the chemical potential. If the origin of energy is shifted to the middle of 
conduction band we have to replace $\varepsilon\to\varepsilon -\mu$, and the 
condition of Cooper instability (\ref{Cupper}) leads to the following equation 
for $T_c$:
\begin{equation}
1=\frac{|U|}{2}\int_{-\infty}^{\infty}d\varepsilon \tilde N_0(\varepsilon)
\frac{th\frac{\varepsilon -\mu}{2T_c}}{\varepsilon -\mu} ,
\label{BCS}
\end{equation}
The chemical potential of the system at different values of $U$ and $\Delta$ now should be determined
from DMFT+$\Sigma$ calculations, i.e. from the standard equation for the number of electrons (band-filling),
determined by Green's function given by Eq. (\ref{Gk}), which allows us to find $T_c$ for the wide range of
model parameters, including the BCS-BEC crossover and strong coupling regions, as well as for different levels
of disorder. This is the gist of Nozieres -- Schmitt-Rink approximation --- in the weak coupling
region superconducting transition temperature is controlled by the equation for Cooper instability 
(\ref{BCS}), while in the strong coupling limit it is determined by the temperature of Bose -- Einstein 
condensation, which is controlled by chemical potential. Then the joint solution of Eq. (\ref{BCS}) 
and equation for the chemical potential guarantees the correct interpolation for $T_c$ through the 
region of BCS-BEC crossover. In the absence of disorder this combination of Nozieres -- Schmtt-Rink 
approximation with DMFT produces the results for the critical temperature, which, as shown in 
Fig. \ref{fig10a}, are almost quantitatively close to exact results, obtained by direct numerical 
DMFT calculations \cite{Keller01,Toschi04,Koga11,JETP14}, but demands much less numerical efforts.

Eq. (\ref{BCS}) demonstrates, that Cooper instability depends on disorder only through the disorder
dependence of the density of states $\tilde N_0(\varepsilon)$, which is the main statement of Anderson theorem.
Within Nozieres -- Schmitt-Rink approach Eq. (\ref{BCS}) is conserved also in the region of
strong coupling, when the critical temperature is determined by BEC condition for compact Cooper pairs. 
However, the chemical potential $\mu$, entering Eq. (\ref{BCS}), may significantly depend on disorder.
In DMFT+$\Sigma$ approximation this dependence of chemical potential (as well as any other single --
particle characteristic) in the model with semi -- elliptic density of states is only due to disorder
widening of conduction band. In this sence both in BCS -- BEC crossover region and in the strong coupling 
limit a kind of generalized Anderson theorem actually holds and Eq. (\ref{BCS}) leads to universal 
dependence of $T_c$ on disorder, due to the change of $D\to D_{eff}$. Such universality
is fully confirmed by direct numerical calculations of $T_c$ in this model, 
performed in Ref. \cite{JTL14}.

\begin{figure}
\includegraphics[clip=true,width=0.5\textwidth]{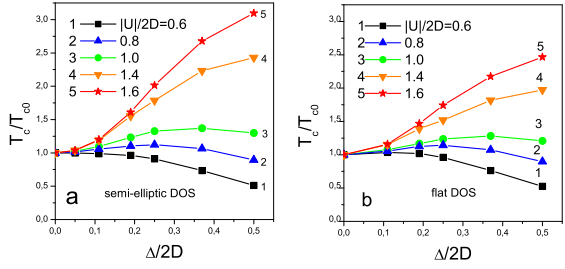}
\caption{Dependence of superconducting transition temperature on disorder for different values of Hubbard
attraction $U$: 
(a) --- semi -- elliptic band.
(b) --- flat band.} 
\label{fig3}
\end{figure}

\begin{figure}
\includegraphics[clip=true,width=0.5\textwidth]{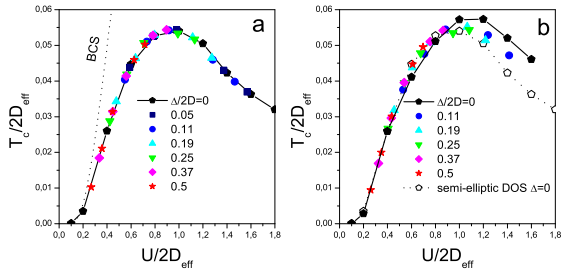}
\caption{Universal dependence of superconducting critical temperature on Hubbard attraction $U$ for different
disorder levels: 
(a) --- semi -- elliptic band. Dashed curve represent BCS dependence in the absence of disorder.
(b) --- flat band. Dashed line represents similar dependence for semi -- elliptic band for $\Delta =0$.}

\label{fig4}
\end{figure}

In Fig. \ref{fig3} we present the dependence of $T_c$ (normalized by the critical temperature 
in the absence of disorder $T_{c0}=T_c(\Delta=0$)) on disorder for different values of pairing 
interaction $U$ for both models of initial semi -- elliptic  density of states (Fig. \ref{fig3}(a)) 
and for the case of flat density of states (Fig. \ref{fig3}(b)). Qualitatively the evolution of 
$T_c$ with disorder is the same for both models. In the weak coupling limit ($U/2D\ll1$) disorder 
slightly suppresses $T_c$ (curves 1). At intermediate couplings ($U/2D\sim 1$) weak disorder 
increases $T_c$, while the further growth of disorder suppresses the critical temperature (curves 3). 
In the strong coupling region ($U/2D\gg 1$) the growth of disorder leads to significant
increase of the critical temperature (curves 4,5). However, this rather complicate dependence of 
$T_c$ on disorder is actually completely determined simply by disorder widening of the initial 
($U=0$) conduction band, demonstrating the validity of the generalized Anderson theorem for all 
values of $U$. In Fig. \ref{fig4} curve with octagons show the dependence
of the critical temperature $T_c/2D$ on coupling strength $U/2D$ in the absence of disorder
($\Delta =0$) for both models of initial conduction bands (semi -- elliptic --- Fig. \ref{fig4}(a) 
and flat --- Fig. \ref{fig4}(b)). In both models in the weak coupling region superconducting 
transition temperature is well described by BCS model (in Fig. \ref{fig4}(a)
the dashed curve represents the result of the solution of BCS model, with $T_c$ determined by 
Eq. (\ref{BCS}), with chemical potential independent of $U$ and determined by quarter -- 
filling of the ``bare'' band), while in the strong coupling region the critical temperature is 
determined by Bose -- Einstein condesation of Cooper pairs and drops as $t^2/U$ with the growth 
of $U$ (inversely proportional to the effective mass of the pair), passing through the maximum -
at $U/2D_{eff}\sim 1$. The other symbols in Fig. \ref{fig4}(a) show the results for $T_c$ obtained 
by combination of DMFT+$\Sigma$ and Nozieres -- Schmitt-Rink approximations for the case of 
semi -- elliptic band. We can see, that all data (expressed in normalized units of $U/2D_{eff}$ 
and $T_c/2D_{eff}$) ideally fit the universal curve, obtained in the absence of disorder. 
For the case of flat band, results of our calculations are shown in Fig. \ref{fig4}(b) and we do not 
observe the complete universality --- data points, corresponding to different degrees of disorder 
slightly deviate from the curve, obtained in the absence of disorder. However, with the growth of 
disorder the flat density of states gradually transforms to semi -- elliptic and our data points 
move towards the universal curve, obtained for semi -- elliptic case and shown by the
dashed curve in Fig. \ref{fig4}(b), confirming the validity of the generalized Anderson theorem
also in this case.

\subsection{Ginzburg -- Landau coefficients}

\begin{figure}
\includegraphics[clip=true,width=0.45\textwidth]{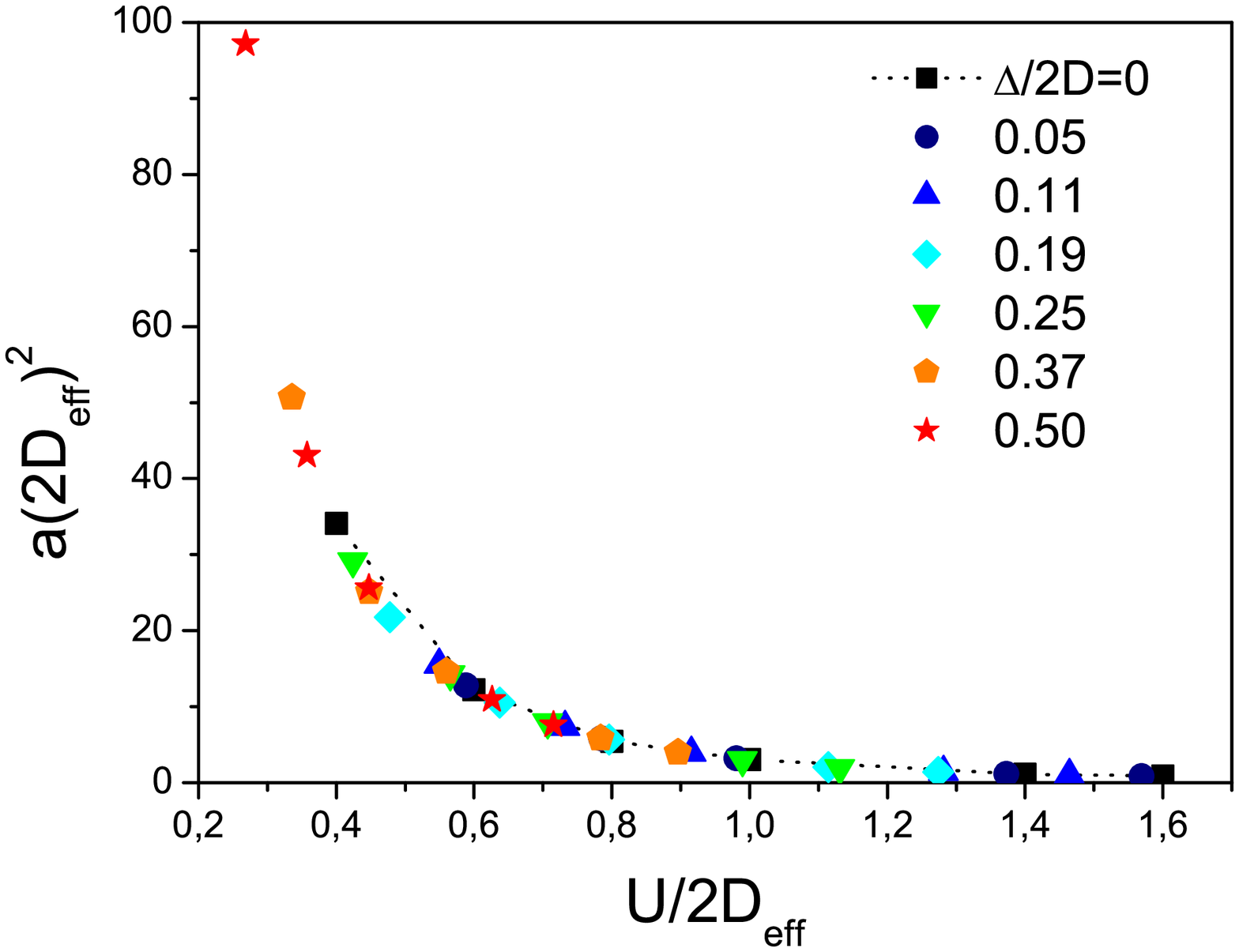}
\includegraphics[clip=true,width=0.45\textwidth]{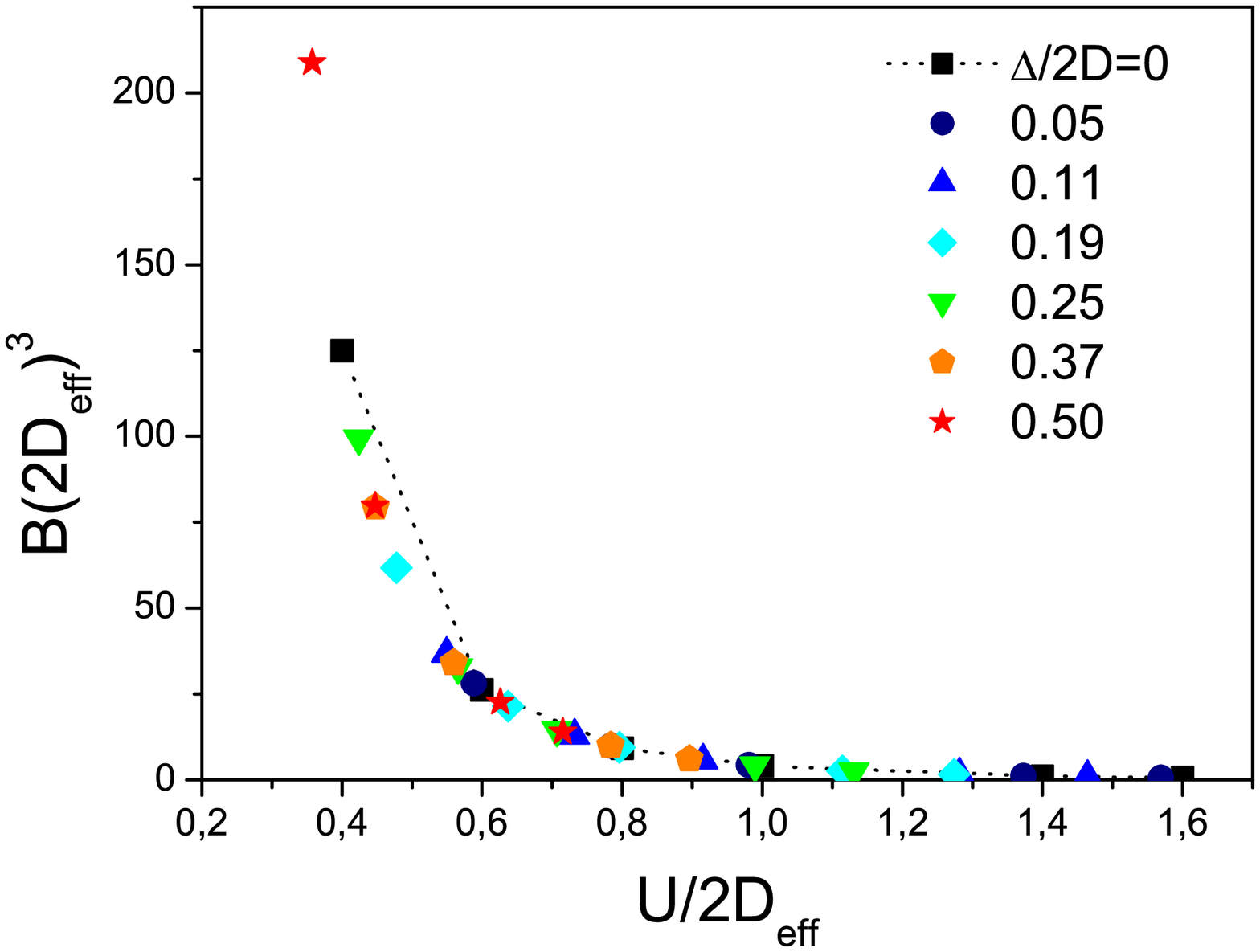}
\caption{Universal dependence of Ginzburg -- Landau coefficients
$a$ (a) and $B$ (b) on Hubbard attraction for different disorder levels.
Dotted line with black squares shows the case of $\Delta =0$.}
\label{fig13}
\end{figure}

Universal dependence on disorder is also observed for the coefficients of
Ginzburg -- Landau expansion $A$ (homogeneous quadratic term of the expansion) 
and $B$ (fourth-order term), related to Cooper -- channel vertices with 
the sum of incoming (outgoing) momenta $q=0$. Coefficient $A$ is given 
by \cite{Diagr}:
\begin{equation}
A(T)=\chi_0(q=0,T)-\chi_0(q=0,T_c),
\label{A_det}
\end{equation}
where $\chi_0(q=0,T)$ is Cooper susceptibility (\ref{chi}), and subtraction of
$\chi_0(q=0,T_c)$ guarantees the zero value of $A(T=T_c)$. 
Using (\ref{Cupper}) to determine $\chi_0(q=0,T_c)$ and (\ref{chi2}) 
for $\chi_0(q=0,T)$, we get:
\begin{equation}
A(T)=\frac{1}{|U|}-
\int_{-\infty}^{\infty}d\varepsilon \tilde N_0(\varepsilon)
\frac{th\frac{\varepsilon -\mu }{2T}}{2(\varepsilon -\mu )}.
\label{A_end}
\end{equation}
so that the coefficient $A(T)$ reduces to zero for $T\to T_c$, and is written 
as:
\begin{equation}
A(T)=a(T-T_c).
\label{A2}
\end{equation}
For the case of ``bare'' band with semi -- elliptic density of states the
dependence of $a$ on disorder is related only to the general widening of
the band by disorder, i.e. is completely described by the replacement 
$D\to D_{eff}$. Thus, in the presence of disorder we obtain the universal
dependence of $a$ on $U$ (normalized by $D_{eff}$), shown in Fig. \ref{fig13}a.

Ginzburg -- Landau coefficient $B$ is determined by the ``loop'' diagramm with
four Cooper vertices \cite{Diagr}.
After rather complicated analysis, to be presented elsewhere, which is based on 
some generalizations of Ward identity (\ref{Word}), it can be shown exactly,
that $B$ is given by:
\begin{equation}
B=\int_{-\infty}^{\infty}\frac{d\varepsilon}{4(\varepsilon -\mu)^3}
\left(th\frac{\varepsilon -\mu}{2T}-\frac{(\varepsilon -\mu)/2T}{ch^2\frac{\varepsilon -\mu}{2T}}\right)
\tilde N_0(\varepsilon)
\label{B_end}
\end{equation}
Thus, the dependence of coefficient $B$ on disorder, similarly to $A$, 
is determined only by the density of states $\tilde N_0(\varepsilon)$ 
renormalized (widened) by disorder and the chemical potential $\mu$. Then, in 
the case of semi -- elliptic density of states the dependence of $B$ 
on disorder is reduced to the simple replacement $D\to D_{eff}$, so that
in the presence of disorder we obtain again the universal dependence of 
$B$ on $U$, shown in Fig. \ref{fig13}b.

It should be noted that Eqs. (\ref{A_end}) and (\ref{B_end}) for coefficients
$A$ and $B$ were otained using the exact Ward identities and remain valid also
in the limit of strong disorder (Anderson localized phase), when both.
$A$ and $B$ depend on disorder also only via the effective bandwidth $D_{eff}$. 

This universal dependence on disorder (due only to the replacement $D\to D_{eff}$) is
reflected also in the specific heat discontinuity at the transitio temperature,
which is determined by coefficients $a$ and $B$:
\begin{equation}
C_s(T_c)-C_n(T_c)=T_c\frac{a^2}{B}
\label{Cs-Cn}
\end{equation}

To determine the coefficient $C$ in gradient term of Ginzburg -- Landau expansion
we need the knowledge of nontrivial of $q$-dependence of Cooper vertex \cite{Diagr}, 
which is essentially changed by disorder scattering. In particular, the behavior
of coefficient $C$ is qualitatively changed at Anderson localization transition \cite{SCLoc}.
Thus, the coefficient $C$ is basically determined by two -- particle charateristics of
the system and does not demopnstrate the universal dependence on disorder due only to
changes of the effective bandwidth.

\subsection{Number of local pairs}

\begin{figure}
\includegraphics[clip=true,width=0.4\textwidth]{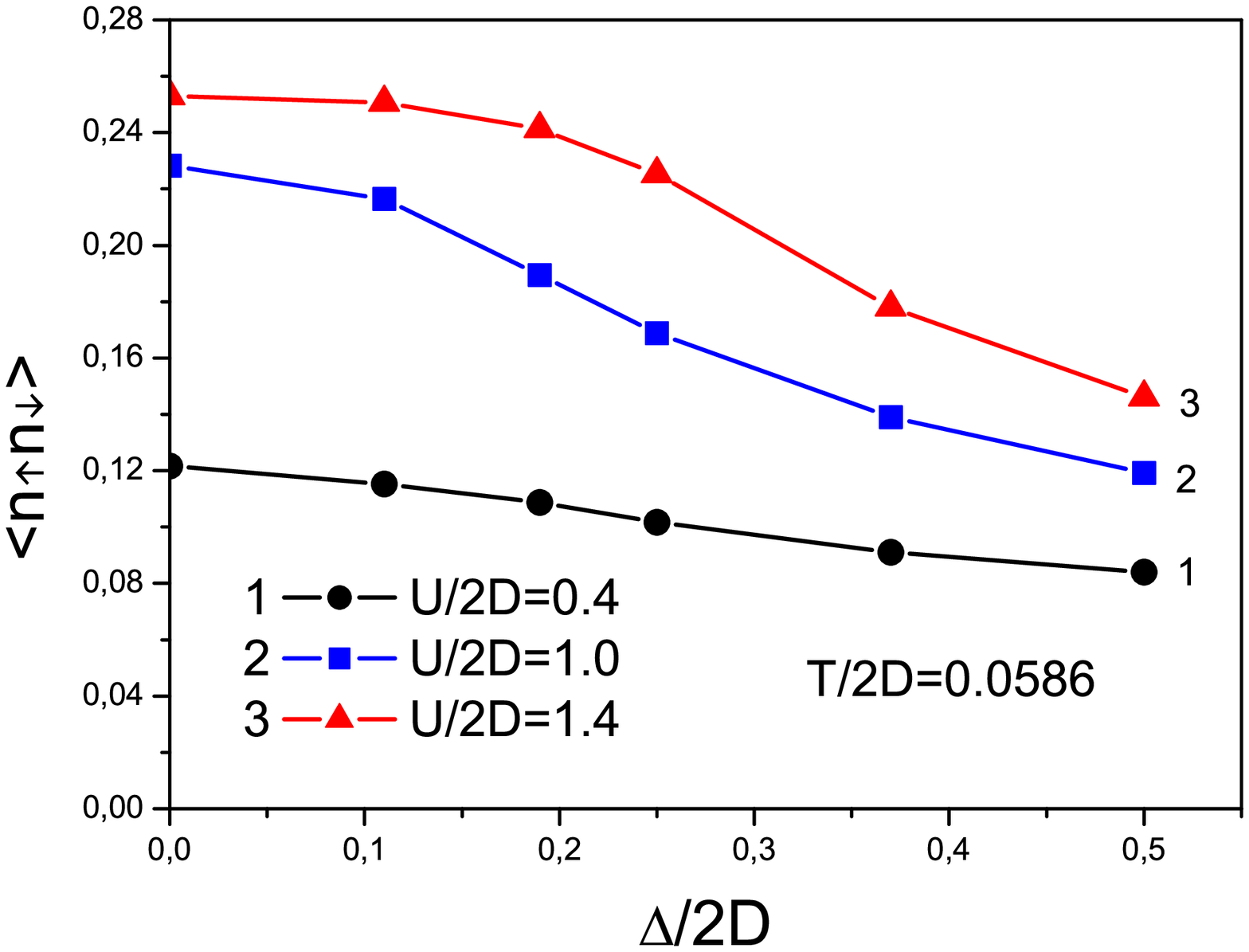}
\includegraphics[clip=true,width=0.4\textwidth]{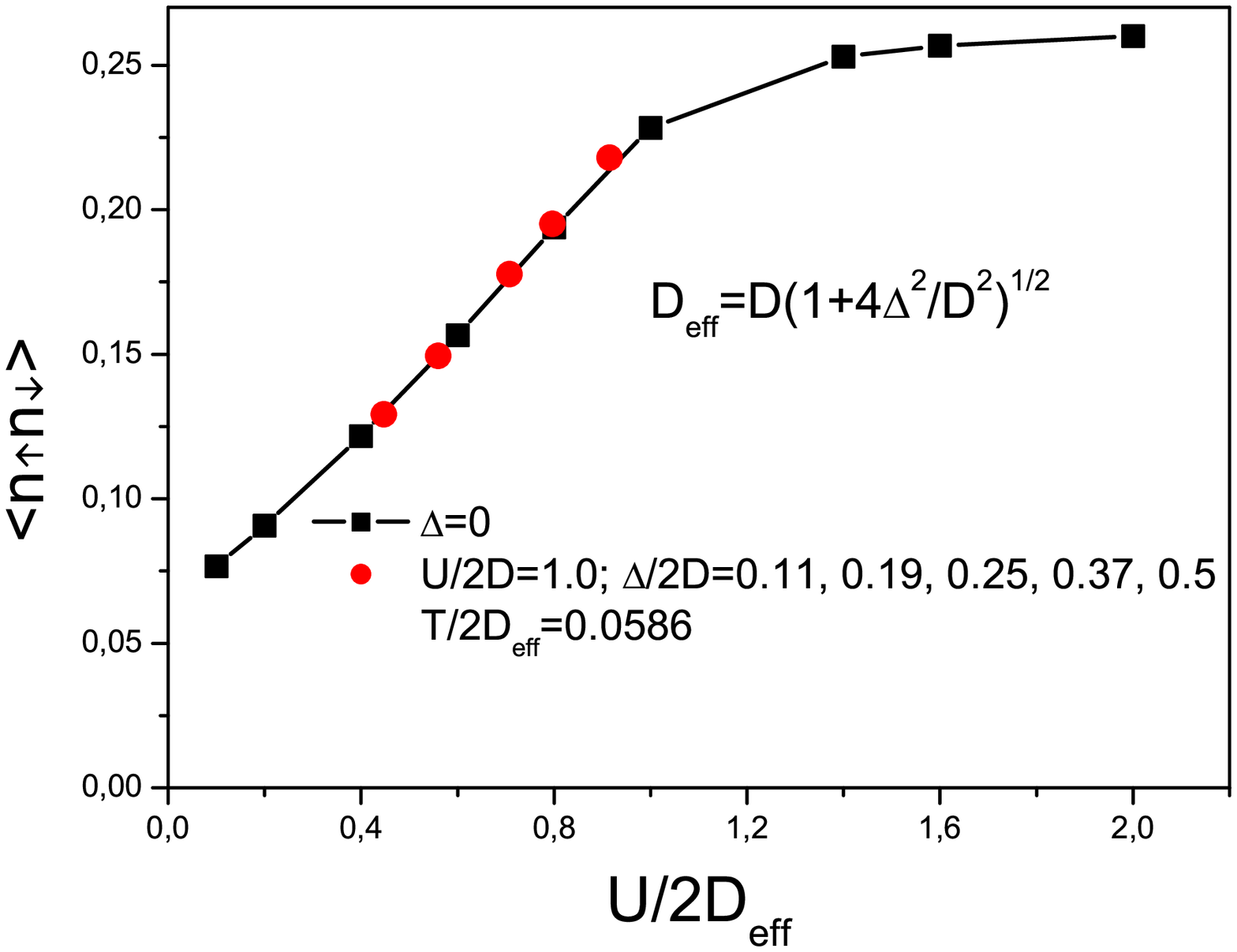}
\caption{Dependence of the number of local pairs on disorder for
different values of Hubbard attraction (a) and universal dependence
on disorder, expressed via normalized variables at fixed value of
$U/2D=1$ (b).}
\label{fig33}
\end{figure}

Disorder in attractive Hubbard model also leads to the suppression of the number of local pairs 
(doubly occupied sites). The average number of local pairs is determined by local (single site) 
pair correlation function $<n_{\uparrow}n_{\downarrow}>$, which in the absence of disorder grows 
with the increase of Hubbard attraction $U$ from 
$<n_{\uparrow}n_{\downarrow}>=<n_{\uparrow}><n_{\uparrow}>=n^2/4$ 
for $U/2D_{eff}\ll 1$ to $<n_{\uparrow}n_{\downarrow}>=n/2$ for $U/2D_{eff}\gg 1$, 
when all electrons become paired. In our calculations $n$=0.5 (quarter -- filled band),
so that $n/2$=0.25, while $n^2/4$=0.0625.
The growth of $D_{eff}$ with disorder leads to an effective suppression of the parameter 
$U/2D_{eff}$ and corresponding suppression of the number of doubly 
occupied sites. In Fig. \ref{fig33}(a) we show the disorder dependence of the number of doubly
occupied sites for three different values of Hubbard attraction. In all cases the growth of 
disorder suppresses the number of doubly occupied sites (local pairs). Similarly to  $T_c$, 
the change of the number of local pairs with disorder can be attributed only to the change of 
the effective bandwidth (\ref{Deff}) with the growth of disorder. In Fig. \ref{fig33}(b) the curve 
with black squares shows the dependence of the number of doubly occupied sites on attractive
interaction in the absence of disorder at temperature $T/2D=0.0586$. This curve is actually 
universal --- the dependence of the number of local pairs $<n_{\uparrow}n_{\downarrow}>$ 
on the scaled parameter $U/2D_{eff}$ with appropriately scaled temperature $T/2D_{eff}=0.0586$ 
in the presence of disorder is given by the same curve, which is shown by by circles 
representing data obtained for five different disorder levels, shown in Fig. \ref{fig33}(b) 
for the case of $U/2D=1$.

\section{Conclusion}

In this paper, in the framework of DMFT+$\Sigma$ generalization of dynamical mean field theory
\cite{UFN12}, we have studied and compared disorder effects in both repulsive and attractive 
Hubbard models. We examined both the problem of Mott -- Hubbard and Anderson metal-insulator 
transitions in repulsive case, and BCS-BEC crossover region of attractive  Hubbard model. 
We also performed extensive calculations of the densities of states and dynamic (optical) 
coductivity for the wide range of ineractions $U$ and at different disorder levels $\Delta$, 
demonstrating similarities and dissimilarities between repulsive and attractive cases.

We have shown analytically for the case of conduction band with semi -- elliptic density of 
states (which is a good approximation for three -- dimensional case) in DMFT+$\Sigma$ 
approximation disorder influences all single -- particle properties (e.g. density of states) 
in a universal way --- all changes of these properties are due only to disorder widening of 
the conduction band. In the model of conduction band with flat density of states (which is 
more appropriate for two -- dimensional systems), there is no such universality in the region 
of weak disorder. However, the main effects are again due to general widening of the band and 
complete universality is restored for high enough disorders, when the density of states 
effectively becomes semi -- elliptic. Similar universal dependences on disorder are also 
reflected in the phase diagram of repulsive Hubbard model and in superconducting critical 
temperature of attractive Hubbard model, where the combination of DMFT+$\Sigma$ and 
Nozieres -- Schmitt-Rink approximations demonstrates the validity of the generalized 
Anderson theorem both in BCS -- BEC crossover and strong coupling regions.

Naturally, no universal dependences on disorder were obtained for the two -- particle properties
like optical conductivity, where vertex corrections due to disorder scattering become very
important, leading to new physics, like that of Anderson transition.

Overall,  the use of DMFT+$\Sigma$ approximation to analyze the disorder effects 
in Hubbard model was shown to produce reasonable results for the phase diagram
for repulsive case, as compared to exact numerical simulations of disorder in DMFT,
density of states behavior and optical conductivity in both repulsive and attractive 
cases. However, the role of approximations made in DMFT+$\Sigma$, such as the neglect of 
the interference of disorder scattering and correlation effects, deserves further studies.

\subsection{Acknowledgements}

It is a pleasure and honour to dedicate this short review to Professor 
Leonid Keldysh 85-th anniversary. This work is supported by RSF grant 
No. 14-12-00502.


\end{document}